\documentclass[preprint,floatfix,showpacs,preprintnumbers,amsmath]{revtex4}

\usepackage{amssymb,graphics}
\usepackage{color}

\begin{document}

\title{An elasto-visco-plastic model for immortal foams or emulsions}
\author{Sylvain B\'ENITO}
\author{Charles-Henri BRUNEAU}
\author{Thierry COLIN}
\affiliation{%
Universit\'e Bordeaux~1,
INRIA Futurs projet MC2 et IMB,
351 Cours de la Lib\'eration,
F--33405 TALENCE cedex, France
}
\author{Cyprien GAY}
\altaffiliation{Universit\'e Paris Diderot--Paris~7
Mati\`ere et Syst\`emes Complexes (CNRS UMR 7057),
B\^atiment Condorcet, Case courrier 7056, 75205 Paris Cedex 13}%
\affiliation{Centre de recherche Paul-Pascal--CNRS, UPR~8641,
Universit\'e de Bordeaux~1,
115 Av. Schweitzer, F--33600 PESSAC, France
}                     
\author{Fran\c{c}ois MOLINO}
\altaffiliation{Centre de recherche Paul-Pascal--CNRS, UPR~8641,
Universit\'e de Bordeaux~1,
115 Av. Schweitzer, F--33600 PESSAC, France
}                     
\affiliation{
Department of Endocrinology,
Institute of Functional Genomics, CNRS, UMR~5203,
INSERM U661, Universities of Montpellier~1 and~2,
141 Rue de la Cardonille, F--34094 MONTPELLIER cedex 05, France
}
%
%

%
\date{\today}

\begin{abstract}
A variety of complex fluids consist in soft, round objects
(foams, emulsions, assemblies of copolymer micelles
or of multilamellar vesicles --- also known as onions).
Their dense packing induces a slight deviation 
from their prefered circular or spherical shape.
As a frustrated assembly of interacting bodies,
such a material evolves from one conformation to another
through a succession of discrete, topological events
driven by finite external forces.
As a result, the material exhibits a finite yield threshold.
The individual objects usually evolve spontaneously
(colloidal diffusion, object coalescence, molecular diffusion),
and the material properties under low or vanishing stress
may alter with time, a phenomenon known as aging.
We neglect such effects to address the simpler behaviour
of (uncommon) immortal fluids:
we construct a minimal, fully tensorial, rheological model,
equivalent to the (scalar) Bingham model.
Importantly, the model consistently describes
the ability of such soft materials
to deform substantially in the elastic regime
(be it compressible or not)
before they undergo (incompressible) plastic creep
--- or viscous flow under even higher stresses.
\end{abstract}

\pacs{
{83.10.Gr}{Constitutive relations}
{83.80.Iz}{Emulsions and foams in Rheology}
{83.50.Ax}{Steady shear flows, viscometric flow}
{83.85.Lq}{Normal stress difference measurements}
     } 

\maketitle

\newcommand{\hs}{\hspace{0.7cm}}
\newcommand{\be}{\begin{equation}}
\newcommand{\ee}{\end{equation}}
\newcommand{\bee}{\begin{eqnarray}}
\newcommand{\eee}{\end{eqnarray}}
\newcommand{\fin}{\nonumber\\}

\newcommand{\upperconv}[1]{^\nabla{#1}}
\newcommand{\lowerconv}[1]{\frac{{\rm D^{(-)}}{#1}}{{\rm D}t}}
\newcommand{\trace}{{\rm tr}}
\newcommand{\transp}[1]{{#1}^{\rm T}}
\newcommand{\transpinv}[1]{{#1}^{\rm -T}}
\newcommand{\inv}[1]{{#1}^{-1}}
\newcommand{\letout}[1]{\widehat{#1}}

\newcommand{\unittensor}{{\rm I}}
\newcommand{\G}{G}
\newcommand{\W}{E}
\newcommand{\mra}{k_1}
\newcommand{\mrb}{k_2}
\newcommand{\energy}{U}
\newcommand{\pdissip}{P_{\rm dissip}}
\newcommand{\screated}{S_{\rm created}}
\newcommand{\moduletensoriel}{{\cal G}}
\newcommand{\stress}{{\cal G}}
\newcommand{\siy}{{\sigma_y}}
\newcommand{\Sis}{{\siy}}
\newcommand{\compl}{{\cal C}}
\newcommand{\defse}{{\cal E}}

\newcommand{\se}{{\epsilon_{\rm s}}}
\newcommand{\dse}{{\rm d}\se}
\newcommand{\Dse}{\dot{\se}}

\newcommand{\vX}{\vec{X}}
\newcommand{\vx}{\vec{x}}
\newcommand{\FF}{F}
\newcommand{\Cauchy}{C}
\newcommand{\LCauchy}{C^-} 
\newcommand{\eLCauchy}{e^-} 
\newcommand{\Finger}{{B}}  
\newcommand{\eFinger}{{e}}
\newcommand{\vxp}{{\vec{x}^\prime}}
\newcommand{\vxpt}{{\vec{x}^{\prime\,{\rm T}}}}
\newcommand{\vv}{\vec{v}}
\newcommand{\gradv}{\nabla\vv}

\newcommand{\si}{\sigma}
\newcommand{\s}{\bar{\sigma}}
\newcommand{\suu}{\s_{11}}
\newcommand{\sud}{\s_{12}}
\newcommand{\sdd}{\s_{22}}
\newcommand{\su}{\s_{1}}
\newcommand{\sd}{\s_{2}}
\newcommand{\strois}{\s_{3}}
\newcommand{\dsi}{{\rm d}\si}
\newcommand{\sip}{\dot{\si}}
\newcommand{\Dsi}{\sip}
\newcommand{\sipl}{\si_{\rm pl}}
\newcommand{\sippl}{\sip_{\rm pl}}

\newcommand{\dt}{{\rm d}t}

\newcommand{\uu}{u}
\newcommand{\tu}{\transp{\uu}}
\newcommand{\U}{U}
\newcommand{\T}{T}
\newcommand{\gd}{\dot{\gamma}}
\newcommand{\est}{{e_{\rm e}}}
\newcommand{\eps}{\varepsilon}
\newcommand{\epsp}{D}
\newcommand{\epspe}{\frac{{\rm D}^+\eFinger}{{\rm D}t}}
\newcommand{\epspp}{{D_{\rm p}^B}}
\newcommand{\epsppt}{\tilde{\epspp}}
\newcommand{\omp}{\dot{\omega}}
\newcommand{\dev}{{\rm dev}}
\newcommand{\rate}{\Gamma}
\newcommand{\ct}{c}
\newcommand{\st}{s}

\newcommand{\hide}[1]{}

\section{Introduction: elasticity and plasticity in foams and emulsions}

\subsection{From crystals to foams and emulsions}

Historically, descriptions of deformations in crystalline, 
solid materials are based on a decomposition 
in terms  of {\em elastic} and {\em plastic} components.
Conceptual and technical problems arise in this process. 
On the one hand, general elastic formulations use 
{\em continous deformations}~\cite{landau_elasticity} 
with respect to a {\em reference state} 
associated with the ordered structure of minimal energy.
On the other hand, plasticity is related to the existence of defects
in the crystalline structure, 
called {\em dislocations}~\cite{landau_elasticity,friedel,kittel},
which are set into motion above some local threshold stress.
Elementary motion steps constitute {\em discrete relaxation events},
which result in a {\em drift} of the reference state.

In foams and dense emulsions, the local arrangement of elementary objects
(droplets, foam cells) can be highly disordered.
In the framework of crystals, this corresponds
to the limit of a very high concentration of dislocations.
Hence, a small increment of stress gives generally access to a large
number of relaxation events. This limits the accessible range
of purely elastic deformations between successive
elementary relaxations.

Like in crystals, each relaxation event is associated with a {\em topological flip}: the
stucture of the network locally changes.
In foams and emulsions, such events are known as ``$T1$ processes'': 
nearest neighbour links are exchanged between two pairs of cells\hide{ (see
Fig.~\ref{t1})}. 
This class of materials thus exhibits an original interplay of elasticity (geometry
and continuity), and plasticity (topology and discreteness).

\subsection{Immortal {\em vs.} aging fluids}

Foams and emulsions usually undergo spontaneous evolution 
such as coarsening (due to coalescence or ripening) 
or drainage~\cite{physics_of_foams,coussot,coussot_english}.
%
Such changes may induce a few topological rearrangements
and cause the material rheological properties 
to slowly evolve~\cite{cipeletti_ramos_2002,cloitre} 
--- a phenomenon known as aging.
If the material is subjected to a weak 
external stress (far below the plastic threshold),
such rearrangements may also statistically induce
some creep which would otherwise 
not occur~\cite{sylvie_cohen_addad_rh_yk_2004}.

In other materials made of soft, round objects,
the relevant molecular processes are slow
and no subtantial aging is observed.
Among such materials --- which may be called
`immortal` --- are copolymer micelles~\cite{molino_fluage1,molino_fluage2}
and some foams and emulsions formulated in such a way
as to make the diffusion of the dispersed phase
and the rate of film ruptures imperceptible
within the experimental time-scale.

\subsection{A brief history of flow localization}
\label{history_localization}

The structural characteristics summarized above
lead to interesting non-linear mechanical behaviours
in which a rich interplay exists
between structural and mechanical responses.
One of the most extensively studied problems
concerns flow localization,
studied in various materials,
from micellar solutions to granular flows.

In the thoroughly studied system
of surfactant solutions self-organized as giant micelles,
the flow curve was originally observed
to exhibit a plateau-like behaviour
under controlled shear rate~\cite{banding_flowcurve}.

Structural observation followed,
demonstrating shear-banding.
Thus, in the stress plateau region, 
two different organizations of the material coexist:
an isotropic region, similar in structure
to the original solution,
and a strongly birefringent region,
in which the micelles are aligned
to a high degree with the flow direction~\cite{banding_birefringence}.

This situation was initially interpretated
in terms of out-of-equilibrium phase transitions
in the material, leading to a steady-state
coexistence between two structurally homogeneous domains~\cite{banding_phasetransition}.

Theoretical descriptions attempted to capture
the onset of this transition and the stability
of the coexistence in terms of interfacial dynamics
and mechanical instability~\cite{banding_model}. 

A more detailed investigation of the birefringent phase
has more recently revealed strong spatial ~\cite{banding_heterogeneous}
and temporal~\cite{becu} variations.
That disagrees with the initial simple picture.

The birefringent phase consists in numerous
transient, narrow zones of very large shear.
The term ``fluid fracture''~\cite{fluid_fracture} has been proposed
to describe these individual events,
which have been observed in different systems
with similar rheological properties
but different internal structures
(connected microemulsions~\cite{molino_microemulsions}, 
copolymer cubic-phase~\cite{molino_fluage1,molino_fluage2}).\newline

The understanding of the shear-banding phenomenology
has thus begun to shift from a phase transition picture
to a fracture picture,
and the interest is now focusing on the highly localized events
that initiate the transition,
and on their relation to the structural properties of the material.

Similarly, in the case of foams, 
a similar phenomenology of localization has been observed 
both in experi\-ments \cite{debregeas_tabuteau_di_meglio}
and simulation~\cite{kabla_debregeas},
and interpreted\footnote{Such interpretations
were initially inspired by reflections
on the dynamics of earthquakes
(individual, local tectonic events
relieve part of the stress locally 
but also report part of it
to other places along the fault~\cite{langer}.}
in terms of the interaction
between individual events
{\em via} (mostly elastic) deformations
of the surrounding material~\cite{falk_langer,picard,picard_2005}.
The challenge now consists in understanding
the self-organization of dispersed relaxation events
into a localized fracture-like behaviour.

Thus, the emergence of a fluid fracture
from these discrete, elementary relaxation events
appears as a well-defined problem.
In that respect, two main problems remain open:
{\em (i)} the role of structural disorder,
and {\em (ii)} non-local effects between individual events,
mediated through elastic stresses in the material
Both problems relate to the unknown typical length scale
on which the discrete system should be averaged
for a descriptions in terms of a continuous model.

More precisely, problem {\em (i)} addresses this length scale
from the limit of smaller length scales
where disorder is relevant.
Conversely, problem {\em (ii)} addresses it 
from the larger length scale limit:
a suitable constitutive equation, 
incroporated into the general framework of continuum mechanics,
provides the tools for generating such non-local effects.

\subsection{Ingredients of our model}

In the present work, we focus on the second problem discussed above,
and construct an example of a rheological model
inspired by the behaviour of such ``immortal'' fluids.
It is characterized by four main features.
\begin{enumerate}
\item The flow properties are motivated and discussed 
in terms of microscopic considerations ($T1$ processes).
\item In order to incorporate the non-local elastic effects
mentioned above, our model is written
in a fully tensorial form,
whether in two and in three dimensions.
\item It is a commonly observed feature that such soft materials
deform substantially before they display plasticity.
In other words, their yield stress
is comparable to their elastic modulus
(unlike for classical, hard crystals).
Correspondingly, the present model implements
a consistent description of the elastic properties
of the material up to {\em finite deformations} ({\it i.e.,}
beyond the usual approximations valid at small deformations).
\item As we shall see now, an amorphous, elastic material
undergoing plasticity loses the memory of past events.
The initial reference state thus has no physical relevance.
Correspondingly, our model is developed in the Eulerian formalism
(attached to the current reference state).
\end{enumerate}


\section{Loss of memory and consequences}


\begin{figure}[h!]
\begin{center}
\resizebox{0.18\columnwidth}{!}{%
  \includegraphics{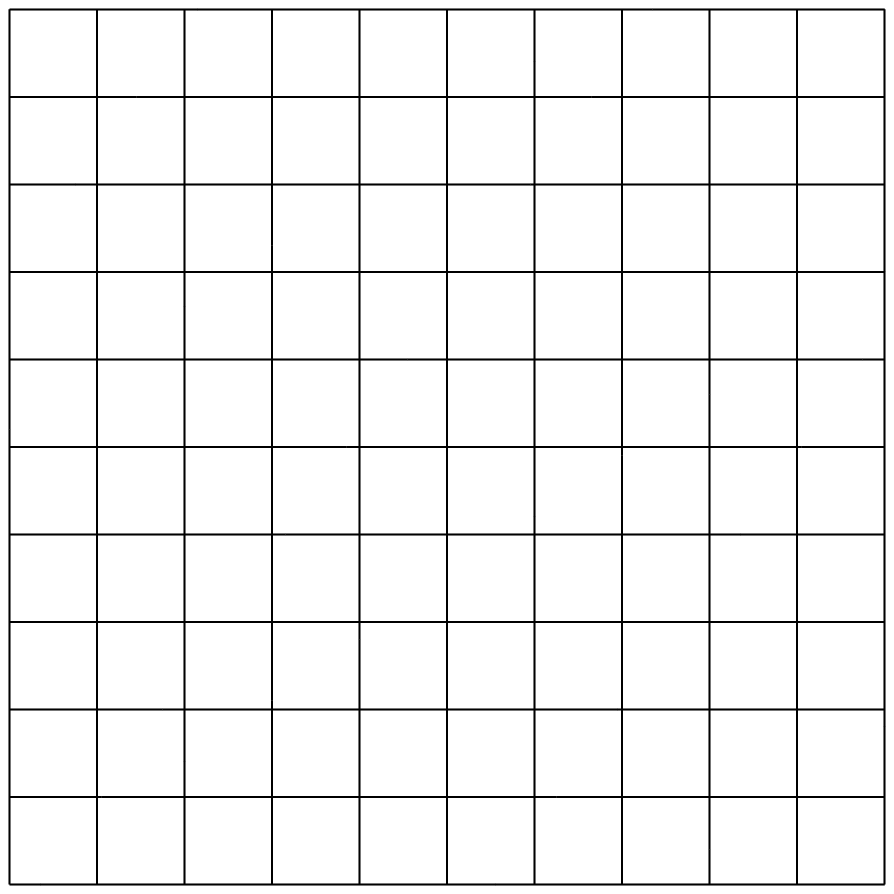}
}
\resizebox{0.18\columnwidth}{!}{%
  \includegraphics{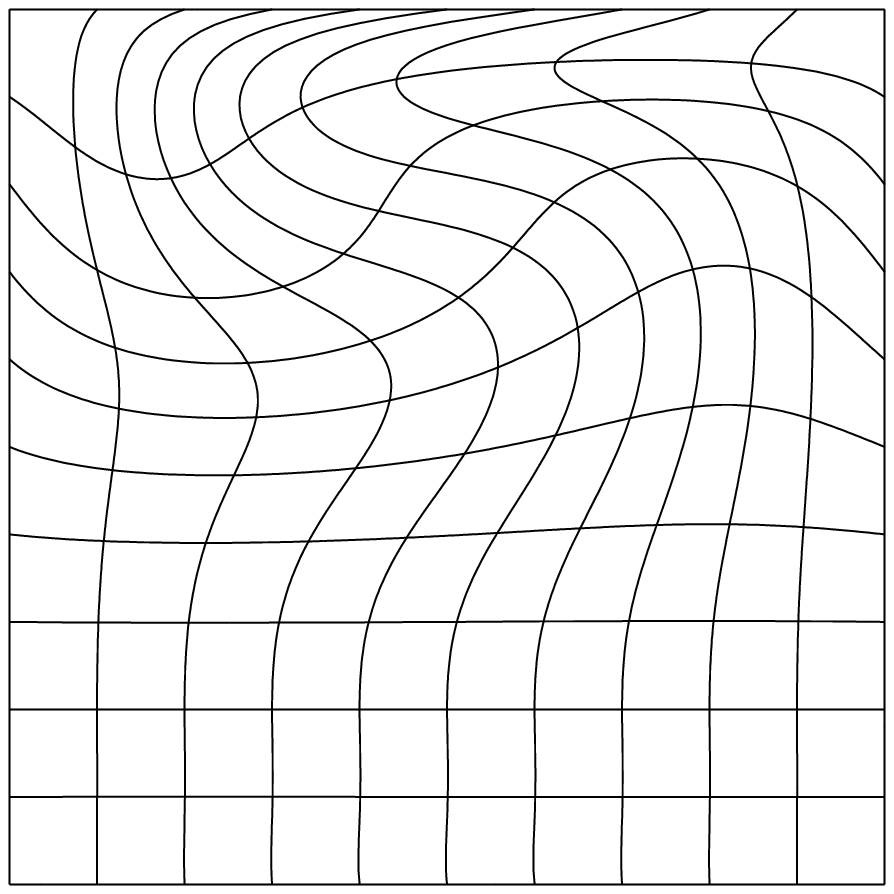}
}
\resizebox{0.18\columnwidth}{!}{%
  \includegraphics{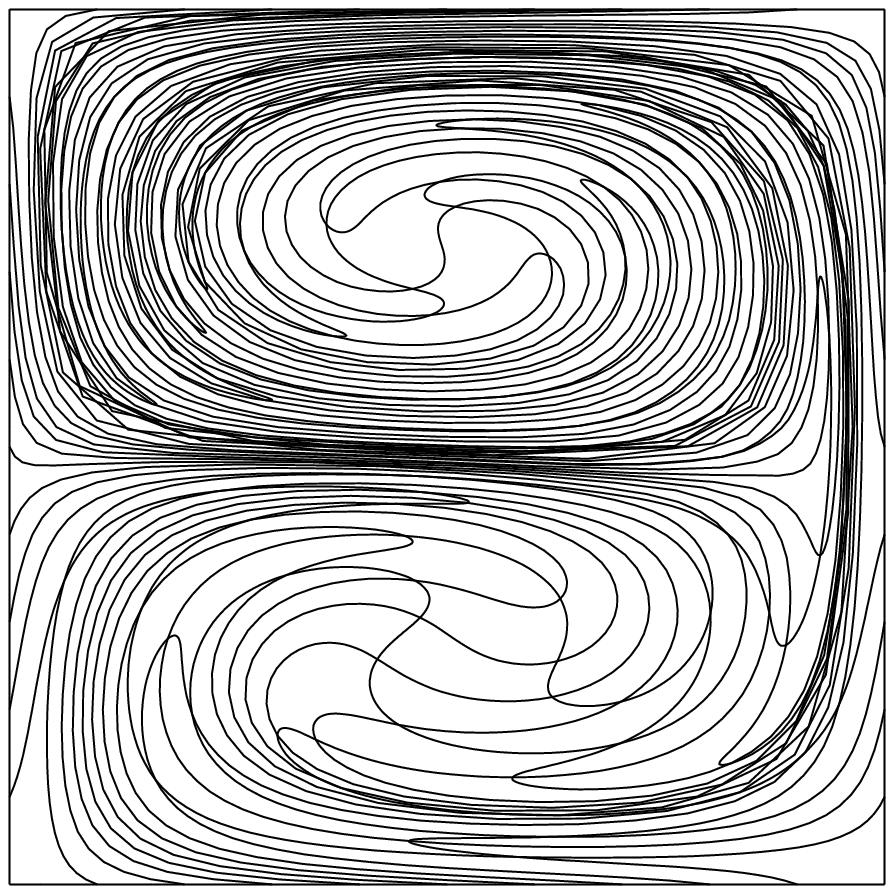}
}
\resizebox{0.2\columnwidth}{!}{\includegraphics{vide.eps}}
\resizebox{0.18\columnwidth}{!}{%
  \includegraphics{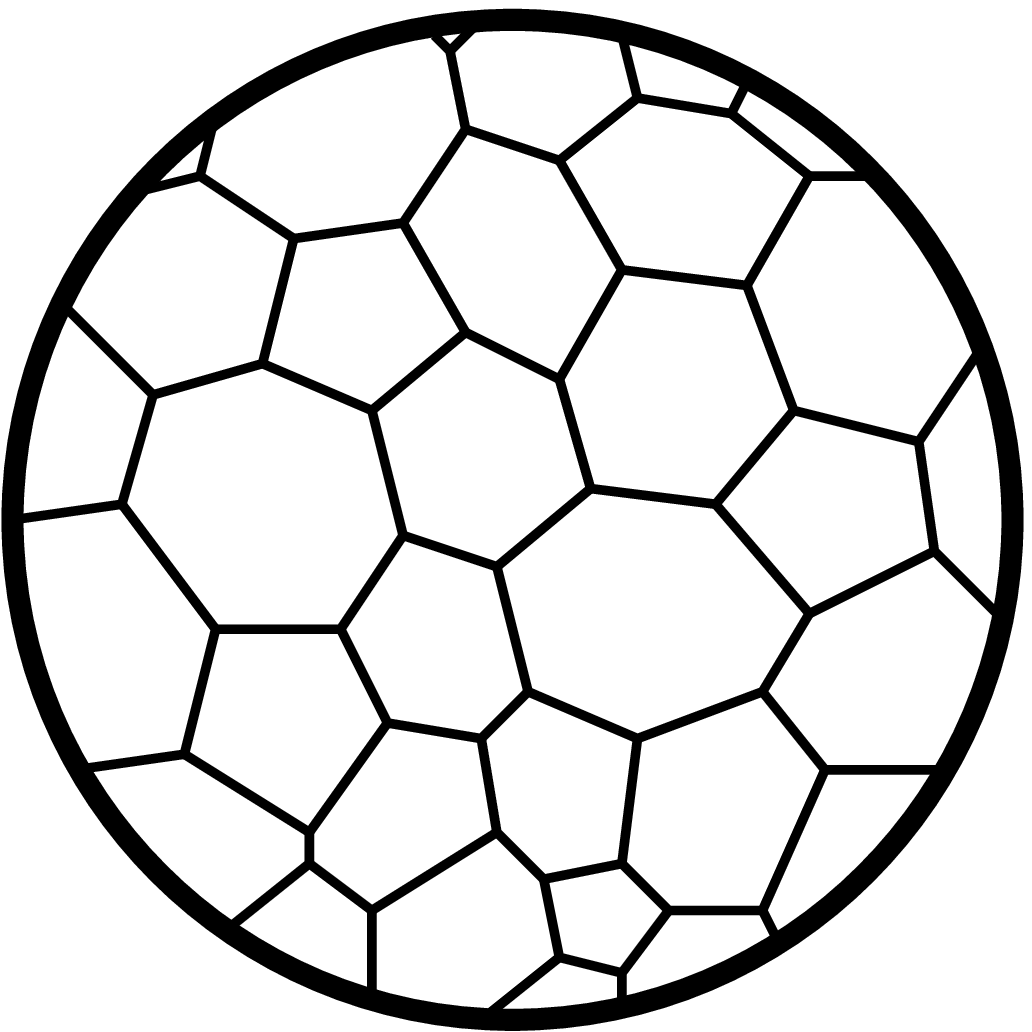}
}
\end{center}
\caption{When a block of foam
undergoes very large deformations
(symbolized by the evolution
of the coordinate system between
the three left-hand side drawings),
one might think that its local structure
correspondingly evolves towards
a highly stretched configuration.
In fact, it always remains
similar to its initial state,
as illustrated on the right-hand side drawing.}
\label{memory_loss}
\end{figure}

In foams or emulsions and in crystals alike,
large deformations of the sample do not imply large deformations
of individual objects, since topological rearrangements relax local
stresses.
In both types of systems, two objects that are initially in contact
can end up at a large mutual distance when many topological
rearrangements have occurred. In such a situation, 
there does not exist any kind of elastic restoring force between both objects.
Hence, the distance between them is irrelevant to the current
mechanical behaviour.
As a consequence, the deformation from the initial state, which keeps
track of such large distances, is mechanically irrelevant, even though
it has a clear experimental meaning.
In other words, the material has lost the memory of such large deformations.

As we shall see, this is the reason why:
\begin{itemize}
\item we use Eulerian coordinates;
\item we define a quantity called ``stored deformation'';
\item we discuss the impact of local disorder.
\end{itemize}

\subsection{Eulerian description}

\begin{figure}[h!]
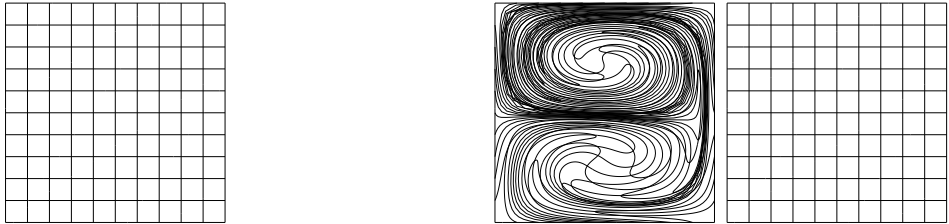

\begin{center}
\resizebox{0.18\columnwidth}{!}{%
  \includegraphics{gdef_init.dat.eps}
}
\resizebox{0.2\columnwidth}{!}{\includegraphics{vide.eps}}
\resizebox{0.18\columnwidth}{!}{%
  \includegraphics{gdef_defmax.dat.eps}
}
\resizebox{0.18\columnwidth}{!}{%
  \includegraphics{gdef_init.dat.eps}
}
\end{center}
\caption{Coordinate systems for a deformed foam.
Let us attach a system of coordinates
to the initial state of a foam (left-hand side drawing).
After some deformation, we may describe the quantities
attached to the foam in terms of either
of two coordinate systems (right-hand side drawings):
either the initial (now deformed) coordinate system,
or a new (undistorted) coordinate system
defined on the current state of the foam.
In the case of elastic deformations,
when material keeps trace of its initial configuration,
such a choice does not have noticeable consequences.
But in the case of deformations that imply
plastic events which progressively erase
the memory of the initial state (see Figure~\ref{memory_loss}),
the choice of the initial, much deformed coordinate system
would not be physically (or computationally!) particularly relevant.}
\label{eulerian}
\end{figure}

In order to describe the deformations of a material sample
and the evolution of the physical quantities attached to it,
two types of coordinate systems are commonly used:
either ``Lagrangian'' coordinate systems attached to the initial state of the sample,
or ``Eulerian'' systems attached to its current configuration,
which coincide to first order in a small deformation context.

In the case of an elastic material,
that keeps the memory of its initial state,
the choice of one or the other
does not have consequences other than computational
(physicists most usually use Lagrangian coordinates~\cite{landau_elasticity}).

However, in the case of a material that progressively looses
the memory of its initial configuration,
such as a foam or an emulsion {\em via} 
rearrangements ($T1$ processes described later in this paper),
it would be physically irrelevant (and computationally tedious)
to refer to the initial sample state.
One therefore generally uses Eulerian coordinates
in such situations.
For instance, the Navier-Stokes equations
are usually expressed in a fixed coordinate system (Eulerian approach).

\subsection{From deformation to `stored deformation'}

Experimentally, the accessible variables are (1) the {\em deformation} or
{\em deformation rate}, as measured or imposed at the sample boundary,
and (2) the {\em stress} (at least in some systems such as foams,
where it can be extracted from the shape of the individual objects, 
or in photoelastic systems).
In order to set up a spatio-temporal numerical scheme,
one needs not only continuity and force balance equations, 
but also a specific evolution relation of the local stress 
in terms of the deformation rate.
Our purpose in this paper is to propose a model example for
this missing ingredient, and to explore its properties.

\begin{figure}[h!]
\begin{center}
\resizebox{1.0\columnwidth}{!}{%
  \input{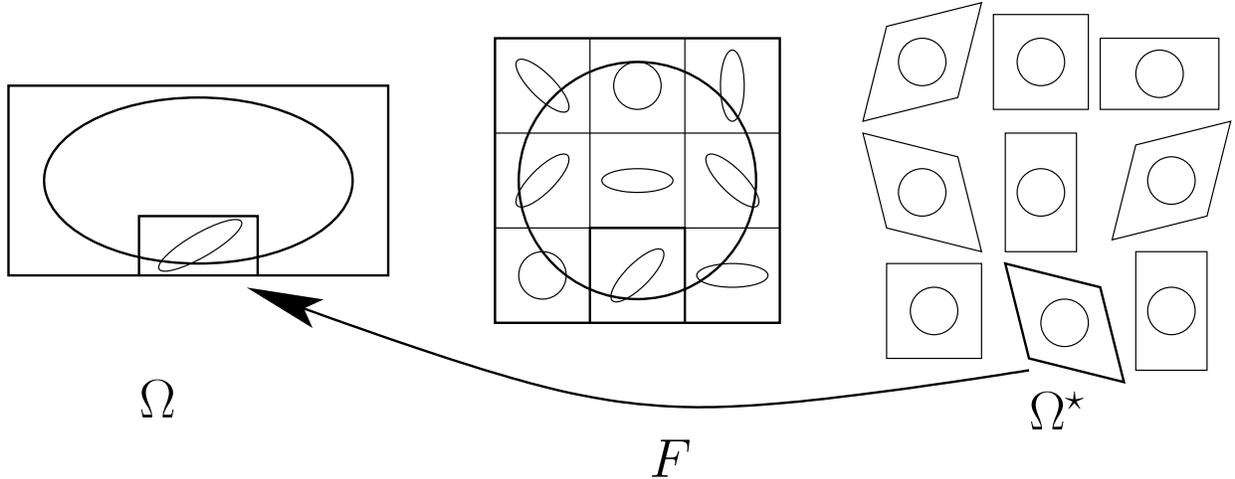}
}
\end{center}
\caption{Stored deformation: operative definition
through a thought experiment.
Consider stretched piece of material $\Omega$,
When its macroscopic deformation
(represented by a large ellipse)
is relaxed (center, large circle),
there remain local internal stresses.
Only by cutting out small pieces of material (right)
can the corresponding stored deformations
(small ellipses) be relaxed (small circles).
The resulting pieces $\Omega^\star$ cannot
fit together without restoring local stresses:
relaxation is meaningful only on the local scale.
Note that the orientation of each piece is arbitrary
and can always be chosen in such a way
that when going back to the initial,
macroscopically stretched state (left),
the corresponding transformation 
($\FF$, see Eq.~\ref{transf_grad_F})
be purely elongational (no rotation),
with stretching factors $\lambda_i$ (see Eq.~\ref{F_lambda_i}).
This stored deformation is related to the local stress
{\em via} the material elasticity (see Eq.~\ref{elasticity}).}
\label{decoupage}
\end{figure}

The very notion of a fixed reference state, and of a global deformation
from this state, being of no use, the stress, as an index of the local
elastic deformation, represents only the {\em recoverable} (or {\em 'stored'}) part of the deformation.
It can be defined through the following thought experiment, described
on Fig.~\ref{decoupage}: a fragment
of the material is cut, in the deformed state, and allowed to relax;
the ``stored deformation'' is defined as the inverse of the
deformation observed during this relaxation.

In the case of foams, it was shown a few years ago~\cite{grenoble_theo_2003}
that it is possible to construct a deformation tensor
from the experimentally observed inter-bubble (cen\-tre-to-cen\-tre) vectors,
which indeed faithfully represents the stress~\cite{grenoble_exp_2003}.

\subsection{Disorder, stored deformation, and stress}

The knowledge of the local stored deformation
is exactly equivalent to the knowledge of the stress. 
They are related through a specific, material-dependent, 
constitutive relation, namely the elasticity, be it linear or not. 

Generically, disordered systems are locally frustrated
and contain internal stresses even in the absence of external applied stress.
In other words, stored deformations are nonzero.
One might think it possible to relax stored deformations
by cutting the material into pieces and sewing them together again, as described above.
In fact, the relaxed pieces do not fit together nicely,
even after adjusting local orientations:
the field of stored deformations
cannot be reconstructed from a displacement field.
If one were to sew all pieces together again,
one would need to stretch each of them appropriately,
thus reconstructing a frustrated stress field
corresponding to a state with zero external applied stress.

The notion of reference state is always clear in a local context.
But it cannot be extended to any macroscopic part of a disordered
material. Indeed, it would not be extensive:
one half of a relaxed sample generally does not match
the relaxed state of the same sample half.

\subsection{Evolution of the stored deformation}
\label{decomp_entrainment_plastic}

Let us return to the construction of the local evolution
of the stress in terms of the applied deformation rate.

The main point is that topological events
participate in the applied deformation
but relax part of the corresponding stored deformation.
The evolution of the stored deformation
must thus involve both an entrainment part and a plastic part.

The entrainment part is purely kinematic, driven by the velocity gradient.
The plastic part reflects the $T1$ relaxation processes
triggered at large stored deformations.
It always\footnote{In the presence
of aging, randomly oriented $T1$ processes 
may sometimes locally {\em enhance} the stored deformation.
On average, however, they are biaised by the ambiant stored deformation
and thus tend to lower it~\cite{sylvie_cohen_addad_rh_yk_2004}.}
tends to lower the stored deformation.
It reflects the rate at which the material
looses memory of the local reference state
which is implicit in the stored deformation.


\section{Choice of a rheological model}
\label{choice_model}

\begin{figure}[h!]
\begin{center}
\resizebox{0.5\columnwidth}{!}{%
  \includegraphics{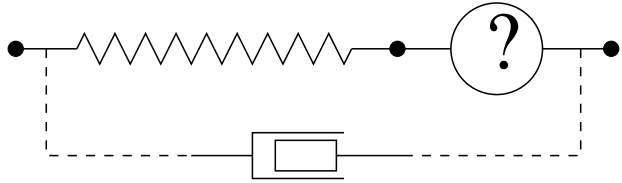}
}
\end{center}
\caption{The considerations of Section~\ref{decomp_entrainment_plastic}
lead us to the idea of a rheological model
consisting in a spring in series with some creeping element
yet to be defined.
The long, soft spring reflects the fact
that creep may trigger at rather large elastic (stored) deformations.
Optionnally, an additional viscous element
may be added in parallel:
in a context of imposed deformation,
it will not alter the dynamics of the system.}
\label{elastic_and_mysterious_creep}
\end{figure}

Let us now choose a rheological model.
The considerations of Section~\ref{decomp_entrainment_plastic}
the rheological model of a foam 
must incorporate a spring (which represents elasticity)
in series with a creeping, plastic part.
In order to choose this plastic part,
let us now review a few common rheological models.
They correspond to the generic form
given by Figure~\ref{elastic_and_mysterious_creep},
where we added an optional viscous element
in parallel with the other two elements.
Note that such a viscous element
impacts the stress response of the material,
but not the local dynamics of stored deformation
(at least not directly\footnote{As the additional stress
that corresponds to the optional viscous element
is transmitted to neighbouring regions of the material,
it may induce additional stored deformation there,
and hence impact the internal dynamics.}).

Such models are listed on Table~\ref{comparison_of_models}
together with an indication of their creep and relaxation properties.
We now review some of these models,
which have been used in the context of foams or similar materials.

\begin{table*}
\label{comparison_of_models}       
\begin{tabular*}{1.0\textwidth}{ccccc}
\hline\noalign{\smallskip}
Model &
& \begin{tabular}{c}Relaxation\\from $\si>\siy$\\towards $\si=\siy$\end{tabular}
& \begin{tabular}{c}Creep\\threshold\end{tabular}
& \begin{tabular}{c}Foams\\(or similar)\end{tabular} \\
\noalign{\smallskip}\hline\noalign{\smallskip}
\resizebox{0.2\textwidth}{!}{\includegraphics{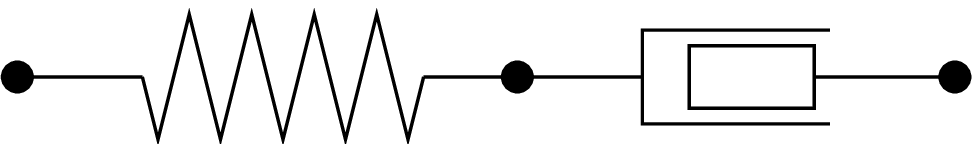}}
& linear Maxwell & delayed 
& \begin{tabular}{c}$\siy=0$\\(viscoelastic)\end{tabular}
& \\
\noalign{\smallskip}\hline\noalign{\smallskip}
\resizebox{0.2\textwidth}{!}{\includegraphics{modele_burger.pstex}}
& linear Burger & delayed 
& \begin{tabular}{c}$\siy=0$\\(viscoelastic)\end{tabular}
& H\"ohler~\cite{sylvie_cohen_addad_rh_yk_2004} \\
\noalign{\smallskip}\hline\noalign{\smallskip}
%
\resizebox{0.2\textwidth}{!}{\includegraphics{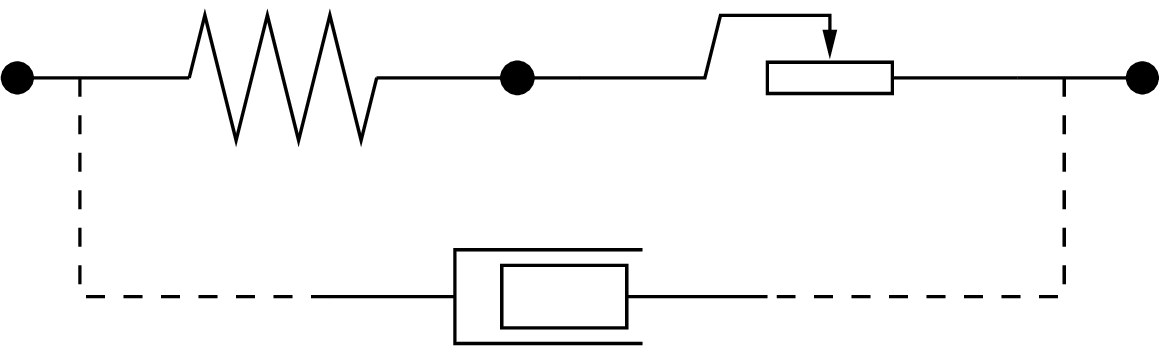}}
& linear elasto-plastic 
& immediate
& $\siy\ll\G$
& \begin{tabular}{c}Marmottant-Graner~\cite{marmottant_graner}\\(+~viscous)\end{tabular} \\
\noalign{\smallskip}\hline\noalign{\smallskip}
%
\resizebox{0.2\textwidth}{!}{\includegraphics{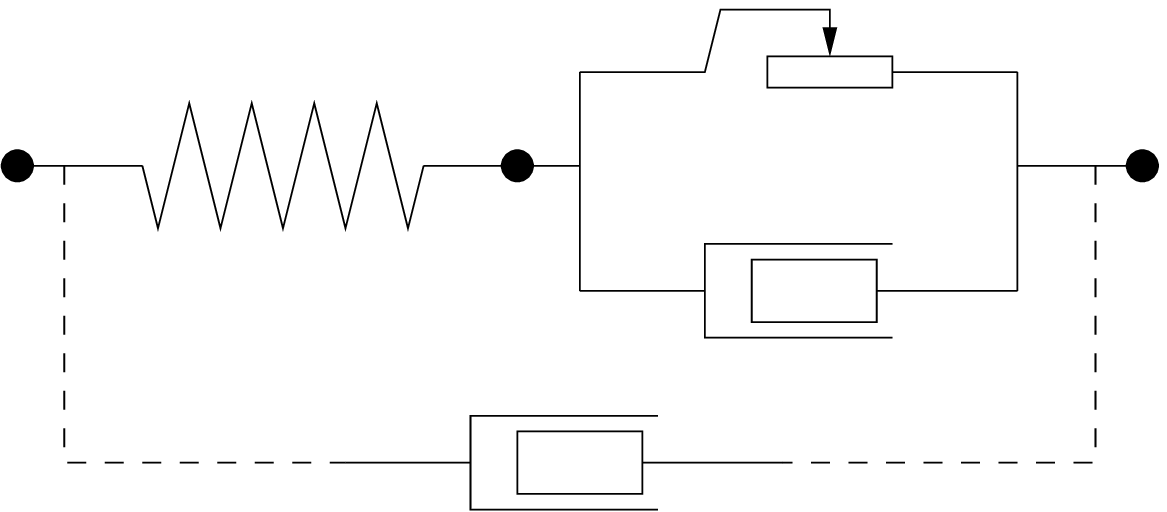}}
& linear Bingham 
& delayed
& $\siy\ll\G$
& \begin{tabular}{c}Saramito~\cite{saramito} (+~viscous)\\Takeshi-Sekimoto~\cite{takeshi_sekimoto}\end{tabular} \\
\noalign{\smallskip}\hline\noalign{\smallskip}
\resizebox{0.2\textwidth}{!}{\includegraphics{ressort_friction_piston_non_lineaire.pstex}}
& non-linear Bingham 
& delayed 
& $\siy\simeq\G$
& present model \\
\noalign{\smallskip}\hline
\end{tabular*}
\caption{Some common rheological models with creep.
Each model is schematically designated
by a combination of springs, viscous and frictional elements,
where viscous elements are not necessarily assumed to respond linearly.
The ability of the material to relax from above the threshold stress
when applied deformation is stopped is indicated.
The value of the stress threshold is compared to that of the elastic modulus
in order to estimate the deformation at the threshold.
Models have been labeled as ``linear'' 
when the deformation at the threshold is small,
whether explicitely or implicitely.
Some references are given when such models have been used
in the context of foams or other soft, disordered materials.
In some instances (marked with label ``+~viscous''), 
an additional viscous element was introduced 
in parallel with the other elements altogether.
This element is then indicated with dashed lines
in the corresponding diagramme in the left-hand side column.}
\end{table*}

\subsection{Burger model for weak applied stresses}
\label{burger_model_for_weak_applied_stresses}

The rheology of dry liquid foams under weak stresses 
is well-described~\cite{sylvie_cohen_addad_rh_yk_2004}
by the Burger model,
which consists in a Maxwell group in series with a Kelvin-Voigt group
(see Table~\ref{comparison_of_models}).
The elastic response (resulting from both spring elements)
corresponds to the deformation of the disordered network
of inter-bubble films and Plateau borders.
The (short) transient (given by the Kelvin-Voigt group)
corresponds to the viscous stretching of films
needed to reach the new equilibrium film conformation.
The (slow) creep (given by the Maxwell viscous element)
corresponds to the spontaneous $T1$ processes 
(which are responsible for aging phenomena)
being slighlty biaised by the ambiant stress.

\subsection{Models with non-zero threshold}

Although well suited to describe the response 
of foams at weak stresses,
the Burger model does not incorporate
the finite plasticity threshold.
Let us now review some models
that do incorporate the threshold,
even though they oversimplify
the short-time response
under weak stresses.

More precisely, these models are designed
to provide most, if not all, of the following features:
{\em (i)} a simple elastic response
to small stresses, 
{\em (ii)} yielding above a stress threshold,
and {\em (iii)} a viscous response 
at large, constant deformation rates.

The first family of such models
includes the simple elasto-plastic model
(a spring in series with a solid friction element),
with an optional viscous element in parallel
(Table~\ref{comparison_of_models}).
The simple elasto-plastic model
does not provide feature~{\em (iii)}.
The viscous element was incorporated
by Marmottant and Graner~\cite{marmottant_graner}
to account for this feature, observed in foams.

The second family, also used in the context of foams
or rheologically similar materials
(see Table~\ref{comparison_of_models}),
includes a viscous element coupled to the solid friction element,
{\em i.e.}, the Bingham model 
(Takeshi and Sekimoto~\cite{takeshi_sekimoto}, 
and present work),
also with an optional viscous element in parallel (Saramito~\cite{saramito}).

In all cases except the pure elasto-plastic model,
the viscous element provides feature~{\em (iii)}.

\subsection{Each foam has its own rheological model}
\label{each_foam_rheological_model}

The reason why different models have been suggested
is that the rheology of a foam varies
with several parameters,
among which surface tension, bubble size,
polydispersity, surfactant properties.
Let us concentrate on the effect of volume fraction
and continuous phase viscosity,
see Figure~\ref{comparaison_mousses_fraction_vol_viscosite_schemas_modeles}.

\subsubsection{Elasticity and volume fraction}

At gas volume fractions below the close-packing threshold,
the foam flows and displays no elasticity.

Slightly above the threshold,
elasticity is weak at small stored deformations
as bubbles move rather freely between neighbouring bubbles,
and strengthens at larger stored deformations
as they come into closer contact with neighbours.

At large gas volume fractions,
the elastic modulus is expected to be large
even at small deformations,
as the bubbles are already in close contact.

\subsubsection{Plastic threshold and volume fraction}

In order for neighbouring bubbles to undergo
a plastic rearrangement (such as a $T1$ process),
they need to deform more importantly
when the gas volume fraction is larger.
As a result, the plastic threshold $\siy$
is also expected to be larger.

\subsubsection{Large elastic deformations}

The plasticity threshold of foams
usually corresponds to moderate deformations,
that are beyond the small deformation regime.
For instance, the threshold deformation
for a polydisperse foam under shear in the dry limit
(volume fraction approaching unity)
is on the order of $30$ to $50\%$~\cite{kraynik_reinelt_polydisperse_foam}.

The ratio $\siy/\G$
is an indication of whether elastic non-linearities
appear prior to the onset of plasticity.
It is not clear to us whether this ratio
increases or decreases with gas volume fraction.

\subsubsection{Relaxation and volume fraction}

Under stationary conditions, 
elasticity is inactive
and both viscous elements play similar roles.
They can be distinguished, however, in transient responses.

At moderate volume fractions,
we expect any relative motion of bubbles
to generate viscous dissipation:
the general viscous element should
dominate over the viscous friction element.

At higher gas volume fractions,
bubbles interact so intimately
that most viscous dissipation
can be expected to arise during plastic events.
In other words, the viscous friction element
should dominate over the general viscous element.

\begin{figure}[h!]
\begin{center}
\resizebox{1.0\columnwidth}{!}{%
  \input{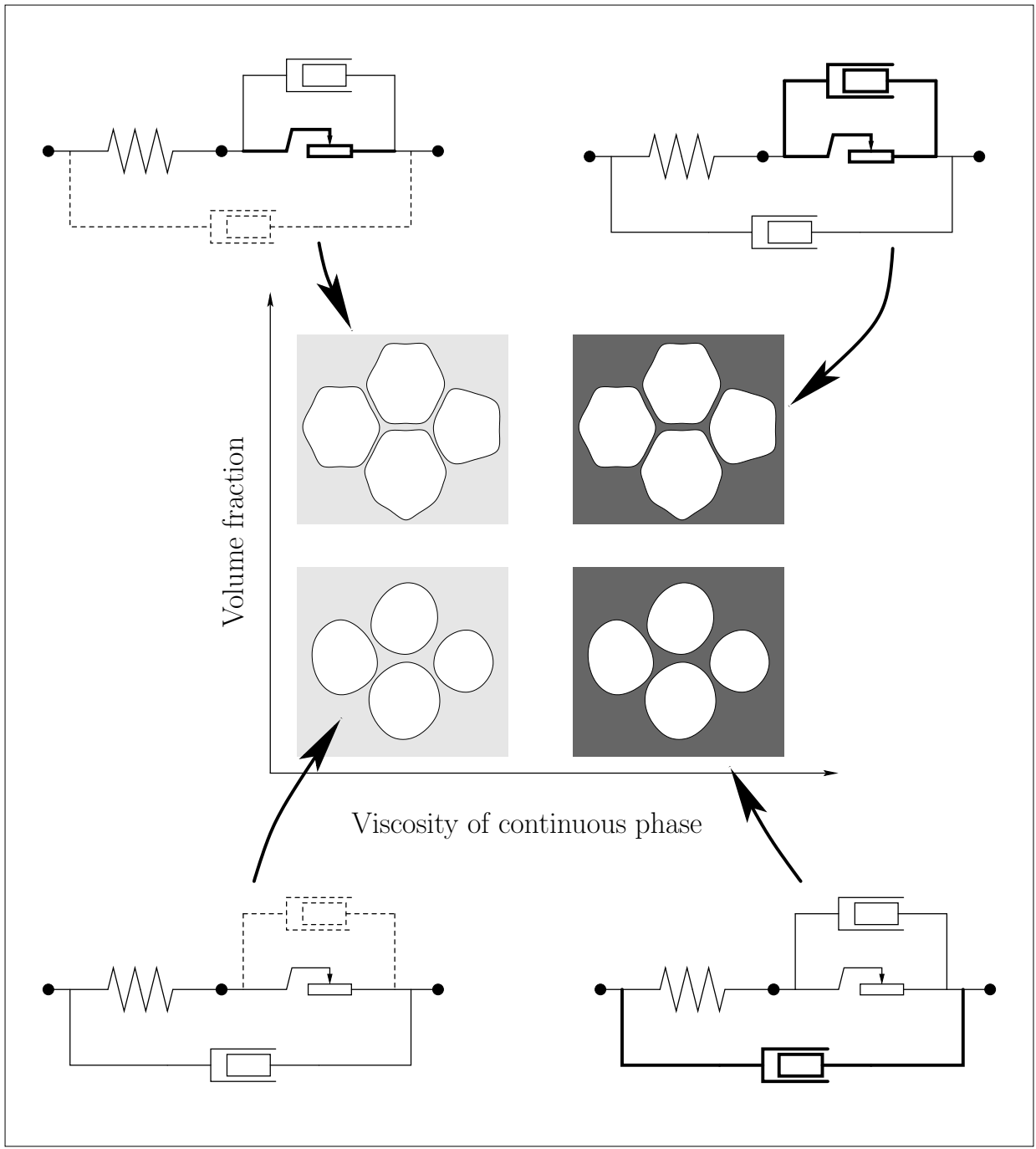}
}
\end{center}
\caption{Tentative variation of foam rheology
as a function of its volume fraction and of the continuous phase viscosity.
Four instances of the local aspect of the foam are drawn 
in the center of the figure,
for low (light grey) or high (dark grey) continuous phase viscosity
and for dispersed (round) or concentrated (faceted) bubbles.
In each case, a tentative corresponding rheological model
is schematically represented in terms of one spring, 
one solid friction element,
one viscous friction element,
and one general viscous element.
The strength of each element is coded as
weak (dashed line), medium (thin solid line) or strong (thick line).
}
\label{comparaison_mousses_fraction_vol_viscosite_schemas_modeles}
\end{figure}

The relative weight of the viscous friction element
and the general viscous element is apparent 
on Figure~\ref{comparaison_mousses_fraction_vol_viscosite_schemas_modeles}.

Their ratio also impacts the ability of the material
to relax when the applied deformation rate
is suddenly brought to zero.
Indeed, in this respect,
only the viscous friction element is relevant.
A model without such a viscous friction element~\cite{marmottant_graner}
does not display relaxation
in common situations (like oscillatory measurements)
where the deformation rate may reverse.

\subsubsection{Viscous {\em vs.} plastic behaviour}

Besides, the visosity of the continuous phase
impacts the relative importance
of the viscous elements and the solid friction element.
This impacts the stationary response of a foam,
which typically changes from mainly plastic
to mainly viscous as the deformation rate is increased.
This transition is expected to occur 
at lower deformation rates
if the continuous phase viscosity is increased.

\subsection{Choice of the Bingham model}
\label{choice bingham}

Let us now choose a specific model
in order to develop a fully tensorial version of it.

Except for sollicitations at very low deformation rates,
the viscous elements impact the rheological response.
Should we keep them both?

The general viscous element provides 
an additional contribution to the stress.
It does not represent any technical difficulty.
Besides, it does not impact the behaviour
of the system under imposed applied deformation conditions.

By contrast, we believe that the viscous friction element
provides essential features such as 
the ability to display relaxation.

For these reasons, in the remaining part of this paper,
we focus on the Bingham model
(see Fig.~\ref{ressort_friction_piston}),
which is the simplest one to provide
all three desired properties
together with relaxation.
We hope that it may also apply
to a broad range of materials made of densely packed,
soft, essentially round objects.

We shall keep in mind
that the parameters of the model
($\G$, $\siy$ and $\eta$ on Fig.~\ref{ressort_friction_piston})
will depend on such physical quantities
as the volume fraction and the viscosity
of the continuous phase, as discussed 
in Section~\ref{each_foam_rheological_model} and illustrated 
by Fig.~\ref{comparaison_mousses_fraction_vol_viscosite_schemas_modeles}.

\subsection{Behaviour of the Bingham model}
\label{behaviour_Bingham_model}

\begin{figure}[ht!]
\begin{center}
\resizebox{0.7\columnwidth}{!}{\input{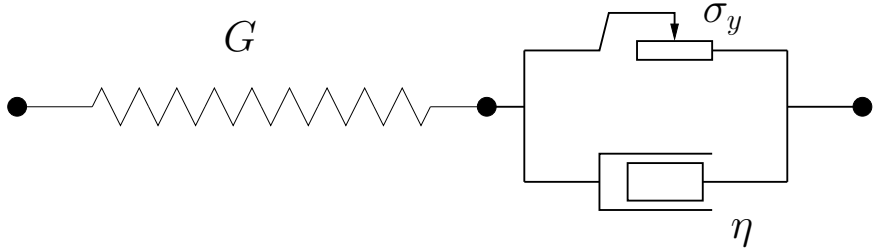}}
\end{center}
\caption{With the Bingham model, from which we develop
a fully tensorial model in the present work,
the response to weak stresses is elastic (modulus $\G$);
by contrast, the response to large stresses 
presents both a yield threshold ($\siy$)
and a viscous component ($\eta$).}
\label{ressort_friction_piston}
\end{figure}

Let us now check (with its scalar version)
that the Bingham model
(Figure~\ref{ressort_friction_piston})
behaves as expected for this class of materials.
In this model, the evolution of stress
(for positive values of the stress\footnote{For real
(positive or negative) values of the stress,
Eq.~(\ref{scalar_elasto_platic_bingham})
must contain an additional term:
$$
\dot{\si}=\G\;\epsp
-\frac{\G}{\eta}\left(
(\si-\siy)\;\theta{[\si-\siy]}
+(\si+\siy)\;\theta{[-\si-\siy]}
\right)
$$}) 
is given by:
\be
\label{scalar_elasto_platic_bingham}
\dot{\si}=\G\;\epsp
-\frac{(\si-\siy)}{\eta/\G}\;\theta{[\si-\siy]}
\ee
where $\theta(x)=1$ when $x\geq0$ and $\theta(x)=0$ otherwise,
and where $\epsp$ is the applied deformation rate,
{\em i.e.}, the symmetric part
of the velocity gradient tensor:
\be
\label{def_epsp}
\epsp=\frac{\gradv+\transp{\gradv}}{2}
\ee

Equivalently, the evolution
of the spring elongation $\varepsilon=\si/\G$
is given by:
\be
\label{scalar_elasto_platic_bingham_e}
\dot{\varepsilon}=\epsp
-\frac{(\varepsilon-\varepsilon_y)}{\eta/\G}\;
\theta{[\varepsilon-\varepsilon_y]}
\ee
where $\varepsilon_y=\siy/\G$.

Let us now consider successively three simple experiments:
(a) quasistatic imposed deformation;
(b) constant, imposed deformation rate;
(c) constant, imposed stress.
The corresponding evolution of the main rheological variables
is schematically represented
on Figure~\ref{trois_manips_sur_le_modele}.

\subsubsection{Quasistatic imposed deformation}

Under low applied deformation, 
the stress depends linearly on deformation.
This low deformation regime is valid as long 
as the resulting stress is smaller than the yield value $\siy$.
At larger deformations, the stress remains constant
and equal to $\siy$, even under arbitrary large 
(but constant) deformations.

The reason why foams and emulsions
display such a solid friction behaviour
is that the $T1$ processes are very similar
to the relaxation of surface bumps
involved in the friction between rough solids:
in both cases, part of the mechanical work done by the imposed stress
is dissipated in discrete relaxation events
which enable discrete deformation steps.
As a result of these events, the work is proportional
to the total deformation (rather than to the velocity,
as in a viscous fluid).

Under imposed deformation, supposing that the deformation value is
reached through a quasi-stationnary process, 
the stress will remain constant as soon as the elastic stress
associated with a deformation increment 
is exactly compensated by the stress relaxed through the plastic processes.
In this situation, even if the imposed deformation can be arbitrarily
large, the {\em stored deformation} remains equal to the value
that corresponds to the yield stress: any extra applied deformation is relaxed
through $T1$ processes.
The noisy aspect of the stress plateau
reflects the disorder of the material
--- and hence, of the distribution of available relaxation processes.

\subsubsection{Constant imposed deformation rate}

Under constant applied deformation rate
(see Figure~\ref{trois_manips_sur_le_modele}b),
the stress rises linearly at short times,
as long as it is smaller than $\siy$.
At later times, it eventually stabilizes above the threshold,
and its final value $\si_\infty$
increases with the deformation rate
(affinely in the Bingham model).

\subsubsection{Quasistatic imposed stress}

\begin{figure}[ht!]
\begin{center}
\resizebox{1.0\columnwidth}{!}{\input{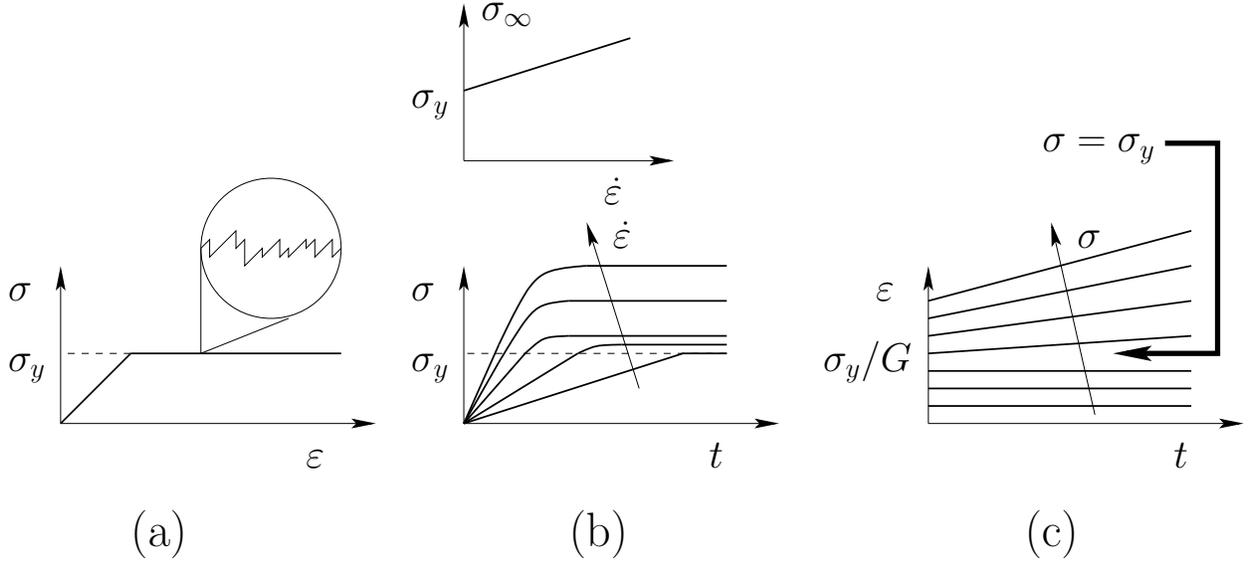}}
\end{center}
\caption{Behaviour of the Bingham model
(time $t$, deformation $\eps$, 
deformation rate $\epsp$, stress $\si$)
in three series of experiments:
(a) quasistatic imposed deformation
(with a zoom on the stress fluctuations
due to individual $T1$ processes
and corresponding elastic loading periods);
(b) constant, imposed deformation rate 
(top: value of plateau stress);
(c) constant, imposed stress.%
}
\label{trois_manips_sur_le_modele}
\end{figure}

Under constant applied stress, by contrast
(see Figure~\ref{trois_manips_sur_le_modele}c),
the system displays two different behaviours.
At small stress values ($\si<\siy$), it behaves elastically.
It switches to a flow behaviour at larger stress values ($\si>\siy$).

\subsection{Two-phase fluid}
\label{two_phase_fluid}

\begin{figure}[ht!]
\begin{center}
\resizebox{1.0\columnwidth}{!}{\input{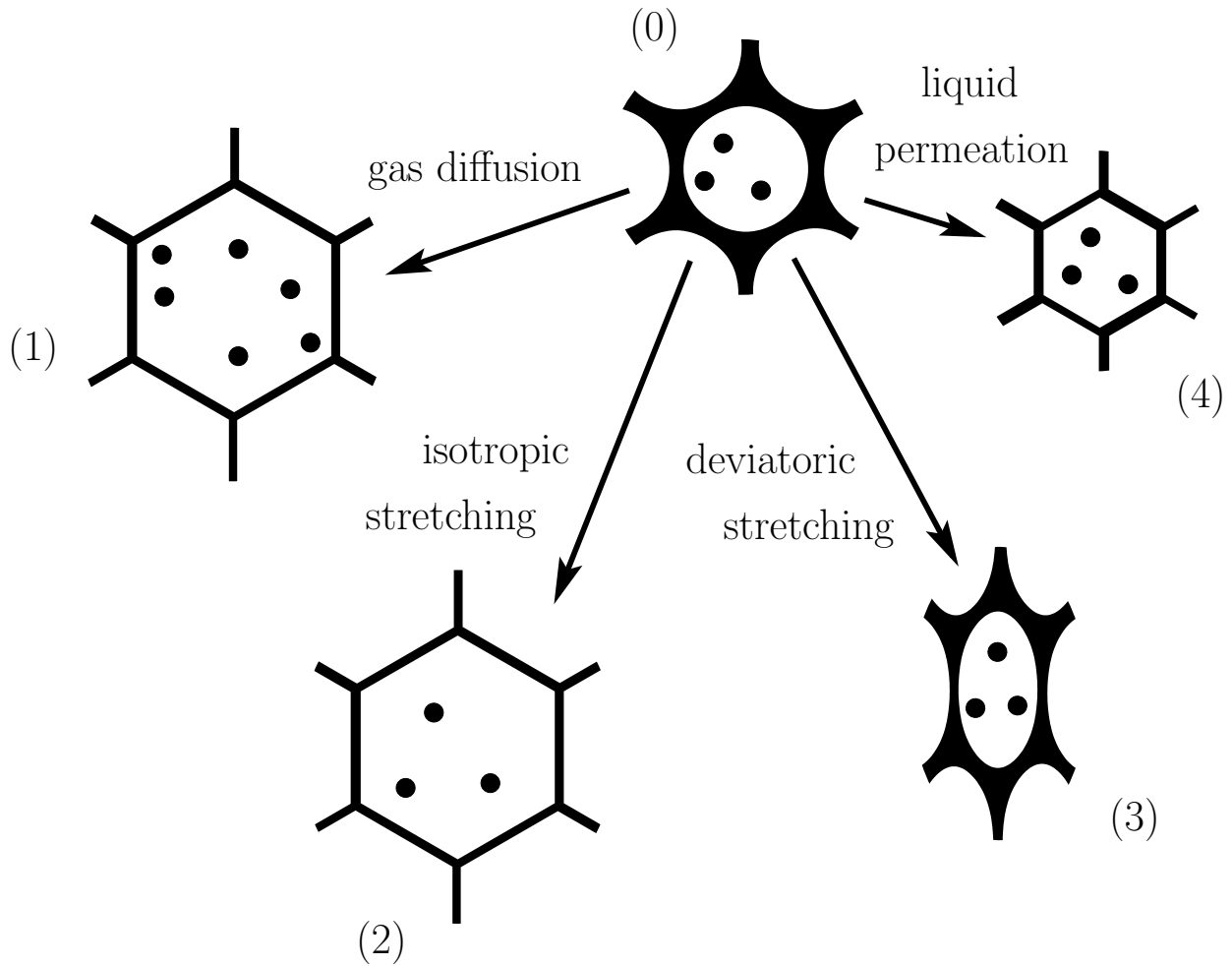}}
\end{center}
\caption{Four changes in the local structure of a foam.
$(0)$ Initial structure. 
$(1)$ Structure obtained through diffusion of gas
from neighbouring regions into the bubble:
the number of gas molecules in the bubble 
(represented by black dots) has increased,
while the amount of liquid that surrounds the bubble has not changed.
$(2)$ Structure obtained by isotropic stretching:
the quantity of gas and the quantity of liquid have not changed.
This is {\em not} a stress-free conformation.
$(3)$ Structure obtained by deviatoric stretching:
the volume, the quantity of gas and the quantity of liquid have not changed.
This is {\em not} a stress-free conformation.
$(4)$ Structure obtained by permeation of liquid
from the vicinity of the bubble towards other places in the foam:
the amount of liquid that surrounds the bubble has decreased,
while the number of gas molecules has not changed.
In the present work, only changes where both phases
are transported {\em simultaneously}, 
such as $(0)\rightarrow(2)$ and $(0)\rightarrow(3)$,
are considered.}
\label{permeation_gas_diffusion_dilation}
\end{figure}

\subsubsection{Evolution modes}
\label{evolution_modes}

Since a foam (or an emulsion)
is a system with two different phases,
it has more deformation modes than a mono\-pha\-sic fluid.
Figure~\ref{permeation_gas_diffusion_dilation}
depicts three isotropic modes obtained from the initial configuration $(0)$
{\em via} gas diffusion $(1)$,
{\em via} an applied isotropic $(2)$ or deviatoric $(3)$ stress,
or {\em via} fluid permeation $(4)$.

Note that the gas diffusion mode
and the liquid permeation mode are plastic,
even for small magnitudes:
they are accompanied by dissipation
and they lead to situations that are stable
even in the absence of any extra applied stress.
Indeed, in situation $(1)$,
the extra amount of gas
occupies the extra volume.
Similarly, in situation $(4)$,
the amount of gas in the bubble 
and the bubble volume have remained constant,
so the force balance within the material
has not been altered despite the loss of liquid.

In the present work, for the sake of simplicity,
we restrict to changes in the applied stress
{\em i.e.}, modes  $(0)\rightarrow(2)$ 
and $(0)\rightarrow(3)$.
For small magnitudes, these two modes are elastic:
if the applied stress is removed,
the system returns to state $(0)$.
For larger amplitudes
mode $(0)\rightarrow(3)$ is plastic.
Indeed, under large applied deviatoric stresses,
rearrangements of the liquid films between bubbles 
(like the $T1$ processes described 
in Section~\ref{T1_threshold} below)
lead to stress-free states
that differ from the initial state,
even though they may locally be very similar,
if not identical, to state $(0)$.

\subsubsection{Density, velocity and stress}
\label{density_velocity_stress}

With this choice of modes $(0)\rightarrow(2)$ 
and $(0)\rightarrow(3)$
rather than modes $(0)\rightarrow(1)$ 
or $(0)\rightarrow(4)$,
the weight fraction of each phase
remains constant, 
and the material can safely
be handled like a one-phase fluid,
which substantially simplifies its description.

The overall material density, $\rho$,
can be expressed in terms of the contributions
from both phases:
\bee
\rho&=&\varphi\,\rho_{\rm liq}+(1-\varphi)\,\rho_{\rm gas}\\
\frac{1}{\rho}&=&\frac{\varphi^w}{\rho_{\rm liq}}
+\frac{1-\varphi^w}{\rho_{\rm gas}}
\eee
where $\varphi$ (resp., $\varphi^w$)
is the volume fraction (resp., weight fraction)
of the liquid phase.


Because we ignore the gas diffusion
and liquid permeation modes depicted
on Fig.~\ref{permeation_gas_diffusion_dilation},
the velocity of the material at larger length scales
suffers no ambiguity, as it is the same 
in the dispersed phase (bubbles or droplets) 
and in the continuous (liquid) phase:
\be
\vec{v}
=\left<\vec{v}_{\rm dispersed\,phase}\right>
=\left<\vec{v}_{\rm continuous\,phase}\right>
\ee

As for the stress,
it varies strongly at the microscopic scale
within such a structured medium as a foam:
compressive within a bubble and across the gas/liquid interfaces,
tensile along these interfaces (surface tension),
tensile within the liquid Plateau borders and vertices.
In the present work, the stress variable $\si$
represents the sum of these contributions
averaged at some larger length scale,
where such structural details are smoothed out:
\be
\si=\left<\si_{\rm gas}\right>
+\left<\si_{\rm gas/liq\,interf.}\right>
+\left<\si_{\rm liq}\right>
\ee

\subsection{Compressibility}
\label{compressibility}

In Section~\ref{choice bingham},
we motivated our choice of the Bingham model
to describe the rheological behaviour
of an emulsion or a foam.
The above considerations, however,
were based essentially on scalar arguments.

One of the main properties that reflect
the actual, tensorial nature
of deformations and stresses in the material
is {\em compressibility},
which is the ability of the material
to adapt its volume when the pressure
({\em i.e.}, the isotropic part of the stress) is changed
(see mode $(0)\rightarrow(2)$ 
in Fig.~\ref{permeation_gas_diffusion_dilation}).
A material is considered {\em incompressible}
when the elastic modulus involved for changes in volume
is much larger than the elastic modulus
involved for deformations at fixed volume.
In practice, the deformations of such a material
therefore obey a fixed volume condition
(mode $(0)\rightarrow(3)$ 
in Fig.~\ref{permeation_gas_diffusion_dilation}).

Among the materials we address in the present work,
emulsions can be considered incompressible
for all practical purposes
since both phases are liquid.

Foams can also be considered incompressible
as long as their constitutive bubbles are not too small.
Indeed, their compression modulus
is then typically equal to the pressure in the dispersed, gas phase,
while their shear modulus
(and other moduli corresponding to deformations at constant volume)
are on the order of the interphase surface tension
divided by the typical bubble radius.
Hence, under atmospheric pressure
and with usual liquids, gases and surfactant molecules,
foams with bubbles not smaller than $0.1~{\rm mm}$ in size
can safely be considered incompressible.
By contrast, the compressibility
of foams made of micron-sized (or even smaller) bubbles
cannot be neglected.

As for bubble monolayers,
when regarded as two-di\-men\-sio\-nal foams,
their apparent compressibility
depends on the boundary conditions.
When such a monolayer is squeezed between two solid plates,
it can be considered incompressible
under the same conditions concerning the typical bubble size
as a three-dimensional foam.
In other situations, a bubble monolayer
may have at least one free interface,
for instance when it floats on a bath of the liquid, continuous phase,
and/or when its upper surface is in contact with the atmosphere.
In such a situation, each bubble is free to slightly deform
in the vertical direction
in order to better accomodate in-plane stresses.
As a result, the bubble monolayer
may appear compressible as seen from above,
even though the total bubble volume
in fact remains essentially constant.

In the present work,
in order to be able to describe the rheology
of all such systems,
we consider a {\em compressible} material
(modes $(0)\rightarrow(2)$ and $(0)\rightarrow(3)$
both available
in Fig.~\ref{permeation_gas_diffusion_dilation}).
Nevertheless, specific properties or mathematical formulations
suitable for the incompressible case
(mode $(0)\rightarrow(3)$ only)
are provided whenever appropriate.


\section{General formulation 
for materials capable of creep}


\subsection{Evolution of the stored deformation}
\label{capable_of_creep}

After the scalar description 
given in Section~\ref{behaviour_Bingham_model}
we now turn to a tensorial version of the model.
In particular, instead of the scalar deformation $\varepsilon$,
we will now use a deformation tensor $\eFinger$,
to be defined in detail
in Section~\ref{relaxed_reference_state_elasticity}.

As mentioned at the beginning of Section~\ref{choice_model},
we are interested in materials 
that display some elasticity and are capable of creep
(see Figure~\ref{elastic_and_mysterious_creep}).
The evolution of their deformation can be decomposed
into an elastic part 
and a creeping part~\cite{these_vincent_mora_2004}.
An example of such a decomposition
was provided above for the (scalar) Bingham model,
see Eq.~(\ref{scalar_elasto_platic_bingham_e}).
More generally, the evolution 
of the stored deformation can be written in the form:
\be
\label{general_decomposition_stored_deformation}
\dot{\eFinger}={\rm kinematics}(\eFinger,\gradv)
-{\rm creep}(\eFinger,\epsp),
\ee
where $\epsp$ is the applied deformation rate
as defined by Eq.~(\ref{def_epsp}).

In the Bingham model, 
the creep term depends only on the stored deformation $e$.
In other models~\cite{marmottant_graner},
it additionally depends on $\epsp$.
This point will be discussed later,
in Section~\ref{section_dissipation}.

\begin{figure}[h!]
\begin{center}
\resizebox{0.5\columnwidth}{!}{%
  \input{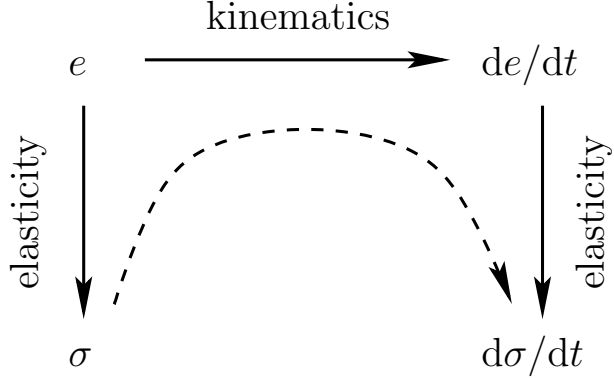}
}
\end{center}
\caption{Evolution of stored deformation $e$ and stress $\si$
in the elastic regime.
The evolution ${\rm d}\eFinger/\dt$
of the stored deformation $\eFinger$ is purely kinematic,
as it directly results from the convection of the material
the velocity gradient $\gradv$.
As for the stress $\si$, it is related
to the current stored deformation $\eFinger$ through the elastic law.
As a result, by composition,
the evolution ${\rm d}\si/\dt$ of the stress
results from the expression of ${\rm d}\eFinger/\dt$
and from the elastic law.}
\label{composition_kinematics_elasticity}
\end{figure}

\subsection{Stress evolution}

For a material capable of creep (see Section~\ref{capable_of_creep}),
once the evolution of the stored deformation is known,
the stress evolves as prescribed by elasticity
(see Figure~\ref{composition_kinematics_elasticity}).

Hence, in the elastic regime,
since stored deformation evolves purely kinematically,
the stress evolution results 
from the kinematics and elasticity alone.
In the presence of plasticity,
although the evolution of the stored deformation
includes an additional term 
(see Eq.~\ref{general_decomposition_stored_deformation}),
the stress evolution is still deduced therefrom 
in the same way, namely {\em via} elasticity.

\subsection{Consistency of some commonly used stress evolutions}
\label{consistency_stress_evolution}

It is common habit to express the stress evolution directly
in the form
\be
\frac{{\rm d}\si}{{\rm d}t}=g(\si,\gradv),
\ee
At first sight, this might seem
exactly equivalent to the evolution for the stored deformation
as given by Eq.~(\ref{general_decomposition_stored_deformation}).

In fact, if no special care is taken,
such an expression does not usually correspond
to the composition of the stored deformation evolution with elasticity
as illustrated on Figure~\ref{composition_kinematics_elasticity}.
The validity of such an evolution equation for stress
is then implicitely restricted to the domain of small stored deformations,
where elasticity is linear.

As an example, as shown in another work~\cite{discussion_on_derivatives},
generalized Maxwell models
that involve the Gordon-Schowalter 
derivative interpolation~\cite{saramito,gordon_schowalter}
suffers such restrictions (except for the special cases
of upper and lower-convected derivatives).

\section{Relaxed, reference state, and elasticity}
\label{relaxed_reference_state_elasticity}

\begin{figure}[ht!]
\begin{center}
\resizebox{0.9\columnwidth}{!}{\input{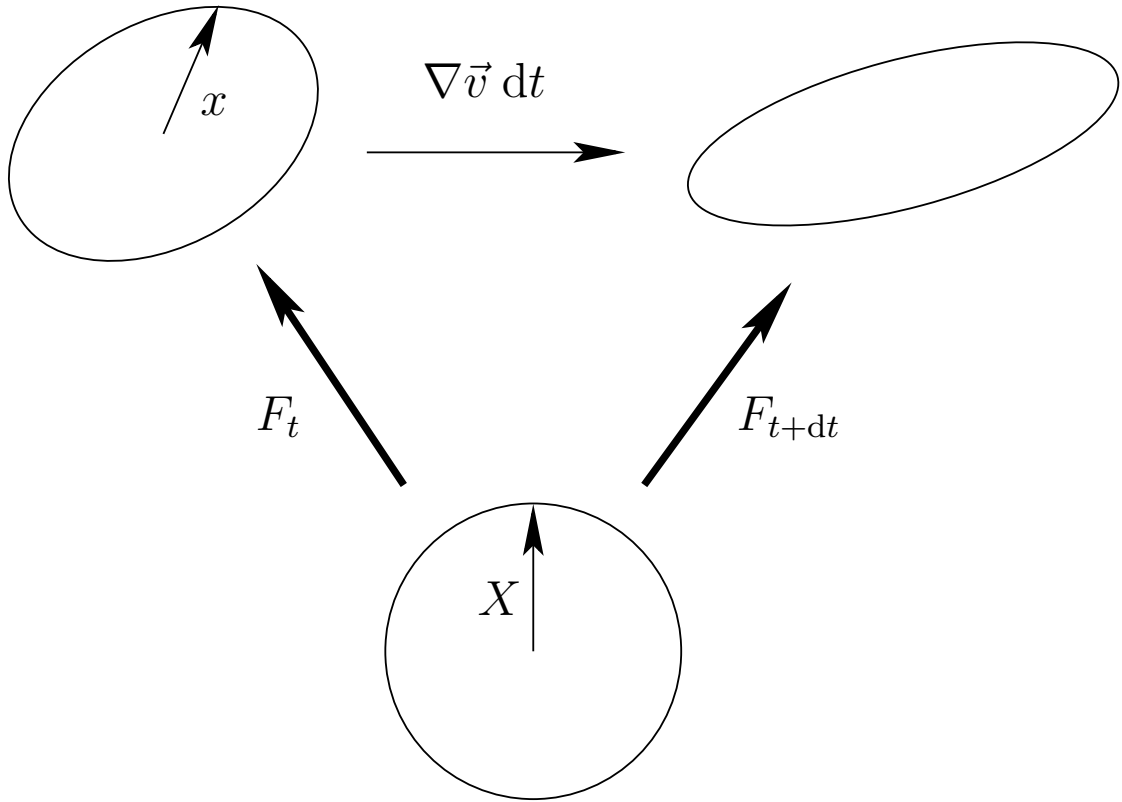}}
\end{center}
\caption{Variation of the local material stored deformation
(see Fig.~\ref{decoupage})
that corresponds to an infinitesimal displacement field.
As described by Eq.~(\ref{transf_grad_F}),
tensor $\FF$ describes the transformation
from the relaxed, reference state (sphere, generic point $\vX$)
to the current state (ellipsoid, generic point $\vx$).
$\FF$ is chosen to be a pure deformation,
while $\gradv\;{\rm d}t$ may include some rotation.
In this example, the evolution of tensor $\FF$
is purely kinematic (see Appendix~\ref{appendix_dBdt}),
as the material is supposed to behave 
in a reversible, elastic manner here.
See Fig.~\ref{dSigma_plus_dEpsilon_avec_prime}
for details on the relaxed state.}
\label{dSigma_plus_dEpsilon}
\end{figure}

Let us now construct a consistent framework 
for the material elasticity,
with the help of Figure~\ref{dSigma_plus_dEpsilon},
which pictures the evolution of the stored deformation
due to an applied flow.

\subsection{Stored deformation}

In such a plastic material, as pointed out above,
there is no point to define the deformation
with respect to some remote, initial reference state.
At any time, however, every region of the material
is stretched (or not).
In other words, its conformation differs (or not)
from the conformation it would have
in the absence of stress from the neighbouring regions:
the stress locally induces ({\em via} its elasticity)
a {\em stored deformation},
which can be visualized in the form of an ellipse 
(or, more generally, of an ellipsoid).
To obtain the ellipse (see Figure~\ref{decoupage}),
one needs to cut out a piece of material,
draw a circle on it while it is thus relaxed,
and put it back in place.

The circle (or, more generally, the sphere)
in the relaxed state is described by:
\be
\label{eq_sphere}
\transp{\vX}\cdot\vX=R^2,
\ee
where $\vX$ is a vector
whose origin is the center of the sphere,
and whose end is a generic point on the sphere
(see Figure~\ref{dSigma_plus_dEpsilon}).
The ellipsoid in the stretched material
can be described by some vector $\vx$,
which depends linearly on $\vX$
if the sphere radius $R$ is infinitesimal,
and can be expressed in terms of the transformation from the relaxed state
to the current state:
\be
\label{transf_grad_F}
\vx=\FF\cdot\vX
\ee
The transformation from $\vX$ to $\vx$
(see Figure~\ref{dSigma_plus_dEpsilon})
can be chosen as a pure deformation (see Fig.~\ref{decoupage}).
Tensor $\FF$ is thus symmetric,
and in its principal axes, we have:
\be
\label{F_lambda_i}
\FF=
\begin{pmatrix}\lambda_1&0&0\\
0&\lambda_2&0\\
0&0&\lambda_3
\end{pmatrix}
\ee
where, in the case of an incompressible material,
$\det\FF=\lambda_1\lambda_2\lambda_3=1$.

The equation for the ellipsoid is obtained
from equations~(\ref{eq_sphere}) 
and~(\ref{transf_grad_F}):
\be
\label{eq_ellipsoid}
\transp{\vx}\cdot\FF^{-2}\cdot\vx=R^2
\ee

\subsection{Finger tensor and associated deformation}

As mentioned in the caption of Figure~\ref{decoupage},
the relaxed state local orientation
can be chosen arbitrarily.
This choice thus must not have any incidence 
on the material elasticity.
We therefore need a variable
that reflects local deformation
without being sensitive to local orientation.
A commonly used such variable
is the Eulerian Finger tensor $\Finger=\FF\cdot\transp{\FF}$.
In the present case where the local orientation is chosen
such that $\FF$ be a pure deformation 
({\em i.e.}, represented by a symmetric tensor),
\be
\label{Finger_beta_i}
\Finger=F^2
=\begin{pmatrix}\beta_1&0&0\\
0&\beta_2&0\\
0&0&\beta_3
\end{pmatrix}
\ee
where $\beta_i=\lambda_i^2$.
From the Finger tensor, we can also construct 
a deformation\footnote{The
deformation $\eFinger$ defined by Eq.~(\ref{eFinger})
happens to coincide, in the present case
where $\FF$ is symmetric (see Fig.~\ref{decoupage}),
with the usual (Lagrangian) deformation
(as defined for instance
by Landau~\cite{landau_elasticity}),
relative to the relaxed state:
$$
e^{\rm Landau}=
\frac12(\transp{\FF}\FF-\unittensor)
$$}
$\eFinger=\frac12(\Finger-\unittensor)$:
\be
\label{eFinger}
\eFinger
=\begin{pmatrix}
\frac12(\beta_1-1)&0&0\\
0&\frac12(\beta_2-1)&0\\
0&0&\frac12(\beta_3-1)
\end{pmatrix}
\ee

\subsection{Evolution of the stored deformation}

Let us now describe how a piece of material under stress
is further deformed when an infinitesimal displacement field
$\vv{\rm d}t$ is applied to the material.
The corresponding deformation,
\be
\gradv\;{\rm d}t,\hs
{\rm where}\hs(\gradv)_{ij}=\frac{\partial v_i}{\partial v_j},
\ee
weakly deforms the ellipsoid (see Figure~\ref{dSigma_plus_dEpsilon}).
As shown in Appendix~\ref{appendix_dBdt},
one can express the equation for the deformed ellipsoid
and derive the evolution equation for tensor $\Finger$:
\be
\label{dFingerdt}
\frac{{\rm d}\Finger}{{\rm d}t}
-\gradv \cdot \Finger
-\Finger \cdot \transp{\gradv}
=0
\ee
where ${\rm d}/{\rm d}t=\partial/\partial t+\vv\cdot\nabla$
is the particulate derivative
and where the entire left-hand side
is the upper-convected objective derivative of $\Finger$.
 
The evolution of the associated deformation, 
defined by Eq.~(\ref{eFinger}),
also involves its upper-convected objective derivative:
\be
\label{dedt}
\frac{{\rm d}\eFinger}{{\rm d}t}
-\gradv \cdot \eFinger
-\eFinger \cdot \transp{\gradv}
= \epsp,
\ee
where $\epsp$ is the applied deformation rate
(see Eq.~\ref{def_epsp}).

Equations~(\ref{dFingerdt}) and~(\ref{dedt})
provide the variation of the stored deformation
in the case of a purely elastic behaviour,
{\em i.e.,} in the absence of any plasticity.
We shall soon discuss the origin of plasticity
and indicate how it may enter such evolution equations.
But let us first give a clear formulation
of how the stress relates to the stored deformation.

\subsection{Elasticity}
\label{disc_elasticity}

Since the material is supposed isotropic,
the principal axes of the stress
coincide with those of the Finger tensor,
and the stress can be expressed
as a linear combination\footnote{This
decomposition results from the {\em Theorem 
of representation of isotropic functions}.}
of three powers of $\Finger$, for instance:
\be
\label{elasticLawIBB2}
\si=a_0\,\unittensor+a_1\,\Finger+a_2\,\Finger^2
\ee
where $a_0$, $a_1$ and $a_2$ are 
scalar, isotropic functions of $\Finger$
({\em i.e.}, of its scalar invariants).
Equivalently, the deviatoric (traceless) part of the stress
\be
\s=\dev(\si)=\si-\frac{\unittensor}{d}\trace(\si)
=\si+p\unittensor
\ee
(where $p$ is the pressure 
and $d$ is the dimension of space)
can be expressed as
\be
\label{s_Finger}
\s=a_1\,\dev(\Finger)+a_2\,\dev(\Finger^2)
=\begin{pmatrix}\su&0&0\\ 0&\sd&0\\ 0&0&\strois
\end{pmatrix}
\ee
and the pressure is given by:
\be
p=-a_0-\frac{a_1}{d}\trace(B)
-\frac{a_2}{d}\trace(B^2)
\ee

More specifically, we will assume that the material
is {\em hyperelastic}, {\em i.e.}, that the stress
results from the differentiation of an elastic potential $\W(\Finger)$,
here defined as being an elastic energy 
per unit mass of the material.
It can then be expressed~\cite{bertram_2005} as:
\bee
\label{sidWdFinger}
\si&=&2\rho\,\frac{{\rm d}\W}{{\rm d}\Finger}\cdot\Finger\\
\label{sdWdFinger}
\s&=&2\rho\,\dev\left(\frac{{\rm d}\W}{{\rm d}\Finger}\cdot\Finger\right)
\eee

with:
\be
\label{dWdFinger}
\rho\;\frac{{\rm d}\W}{{\rm d}\Finger}
=\frac{a_0}{2}\,\Finger^{-1}
+\frac{a_1}{2}\,\unittensor
+\frac{a_2}{2}\,\Finger
\ee

In the case of an incompressible material
(see Section~\ref{compressibility} for examples),
the stored deformation
(expressed in terms of $\Finger$ or $\eFinger$)
does not depend on pressure:
it only depends on the deviatoric
part $\s$ of the stress.
Conversely, the deviatoric stress $\s$
can be expressed in terms of $\Finger$ or $\eFinger$:
\be
\label{elasticity}
\s={\cal F}(\Finger)
={\cal G}(\eFinger)
\ee

In this incompressible limit,
the pressure $p$ (more precisely, its term $a_0$)
varies very strongly
when $\det\Finger$ departs from unity.
In practice, one usually considers
that $p$ is not known explicitely:
in a practical situation,
it must be obtained from the boundary conditions,
while strictly enforcing the constraint $\det\Finger=1$.

\section{Plasticity}

Now that we have described the evolution
of the stored deformation in the material
(Eq.~(\ref{dFingerdt}) or~(\ref{dedt}))
and the material elasticity (Section~\ref{disc_elasticity}),
let us turn to the description of its plastic properties.

As mentioned earlier, a plastic term
must be added to Eq.~(\ref{dFingerdt}) 
or~(\ref{dedt})
in order to reflect the deformation due to topological events:
\be
\label{dFingerplast}
\frac{{\rm d}\Finger}{{\rm d}t}
-\gradv \cdot \Finger
-\Finger \cdot \transp{\gradv}
=-2\,\epspp
\ee
\be
\label{dedtplast}
\frac{{\rm d}\eFinger}{{\rm d}t}
-\gradv \cdot \eFinger
-\eFinger \cdot \transp{\gradv}
= \epsp -\epspp
\ee
This equation is the tensorial version
of the decomposition of the applied deformation
into a kinematic part and a creep part
expressed by Eq.~(\ref{general_decomposition_stored_deformation}).

In the present section, we recall the microscopic origin
and discuss the mathematical properties
of the plastic term $\epspp$ in the evolution equation.

\subsection{Threshold for a single $T1$ process in two dimensions}
\label{T1_threshold}

\begin{figure}[ht!]
\begin{center}
\resizebox{0.9\columnwidth}{!}{\input{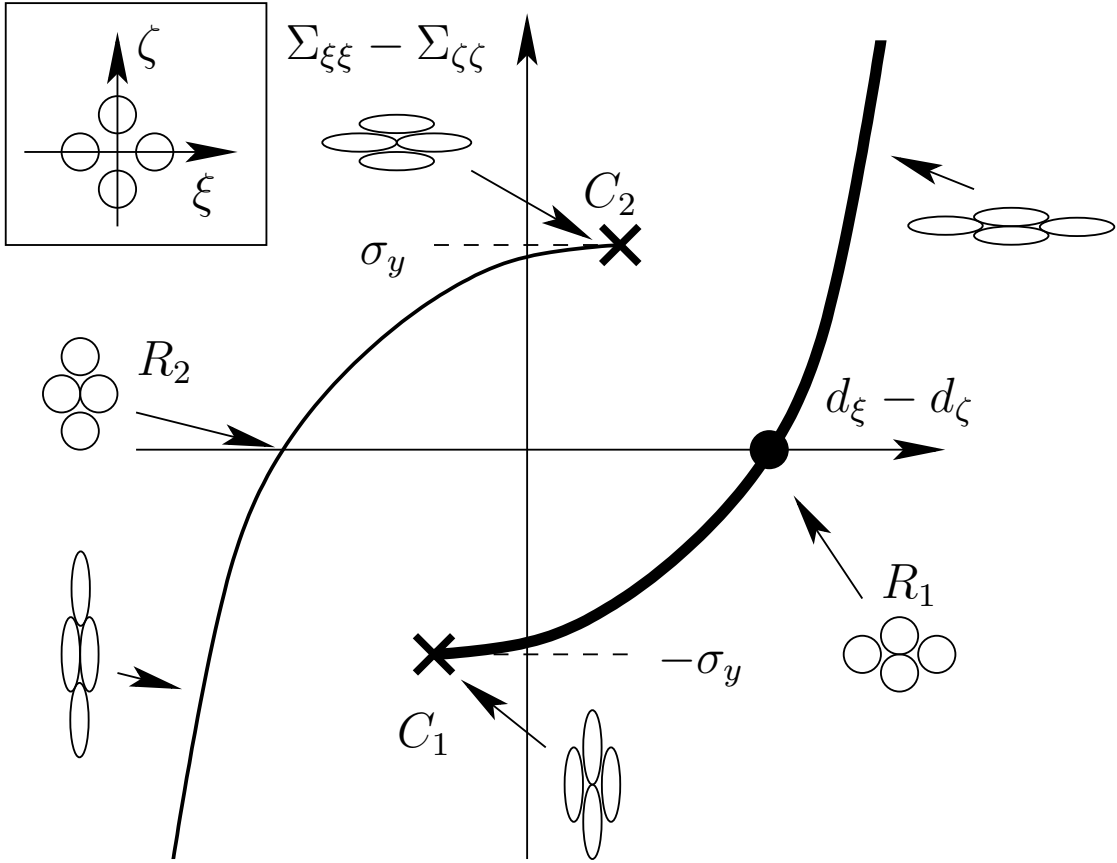}}
\end{center}
\caption{Schematized $T1$ process in two dimensions:
conformations of four neighbouring objects (quadruplet)
subjected to external forces.
One pair of objects is oriented along axis $\xi$
and the other one along axis $\zeta$ (see inset).
The stress principal axes are assumed to be aligned
with those of the quadruplet,
which therefore adopts a diamond configuration.
Moreover, the applied pressure is assumed to be constant.
The diagramme indicates how both pairs compare
as for the (centre-to-centre) 
inter-object distance (horizontal axis)
and as for the external stress 
that they undergo (vertical axis).
By convention, positive values 
of the stress $\Sigma_{\xi\xi}$ (resp. $\Sigma_{\zeta\zeta}$)
denote traction on the pair that is oriented 
along axis $\xi$ (resp. $\zeta$).
A $T1$ process starts from the relaxed situation $R_1$,
denoted by a black circle.
Through compression along axis $\xi$ ({\em i.e.,} $\Sigma_{\xi\xi}<0$)
or traction along $\zeta$ ($\Sigma_{\zeta\zeta}>0$)
or through a combination of both,
the system follows branch~1 (thicker line)
towards the critical point $C_1$.
The $T1$ itself is the sudden transition
from $C_1$ (on branch~1) to branch~2 (thinner line).
In the case of an incompressible material
the quadruplet conformation depends
only on $\Sigma_{\xi\xi}-\Sigma_{\zeta\zeta}$,
while for a compressible material it additionally depends 
on $\Sigma_{\xi\xi}+\Sigma_{\zeta\zeta}=-p$.%
}
\label{t1_df_dx}
\end{figure}

Figure~\ref{t1_df_dx} provides a two-dimensional illustration 
of a $T1$ process in an emulsion or a foam.
Four objects (hereafter called {\em quadruplet})
are initially organized in a diamond conformation.
As compared to the initial conformation at rest (point $R_1$),
a weak applied stress induces an elastic deformation,
symbolized by a continuous branch (thick line).
Once the applied stress exceeds a certain value
(symbolized by $\siy$ on Fig.~\ref{t1_df_dx}),
the system undergoes a discrete flip
while the four objects swap neighbours.
This is the $T1$ process,
symbolized by a jump onto the other branch.

More precisely, let us suppose that the quadruplet
is symmetrical like that on Fig.~\ref{t1_df_dx},
with axes $\xi$ and $\zeta$.
The stress orientation that is most favourable
for the $T1$ process to occur
is that in which the stress axes
are aligned with those of the quadruplet.
Thus, Fig.~\ref{t1_df_dx} is drawn in terms
of $\Sigma_{\xi\xi}$ and $\Sigma_{\zeta\zeta}$,
with the non-diagonal stress component 
$\Sigma_{\xi\zeta}$ being equal to zero.

The condition for the quadruplet to remain on branch~1
of Figure~\ref{t1_df_dx} can be expressed in the form:
\bee
\label{seuil_Sigma}
&&f_{\rm eigen}(\Sigma)\leq 0 \\
{\rm where}&&
\label{seuil_Sigma_2d}
f_{\rm eigen}(\Sigma)=\Sigma_{\zeta\zeta}-\Sigma_{\xi\xi}-\Sis^{T1}
\eee
Here, $\Sis$ is a real, positive number.
When the pressure is varied
or when the stress is not aligned with the quadruplet
(non-zero shear stress component,
{\em i.e.}, $\Sigma_{\xi\zeta}\neq 0$ in the present, 2d case),
it more generally takes the form:
\be
\label{SiST12d}
\Sis^{T1}=\Sis^{T1}(p,\Sigma_{\xi\zeta})
\ee

$\Sis^{T1}$ is expected to increase with the applied pressure
$p=-\Sigma_{\xi\xi}-\Sigma_{\zeta\zeta}$.
Indeed, increasing the pressure reduces the typical object size,
thus enhancing surface tension effects.
It should therefore increase the stress threshold
in such surface tension sensitive systems
as foams or emulsions.

In the case of an incompressible material,
$\Sis^{T1}$ depends only on $\Sigma_{\xi\zeta}$, not on $p$.
More generally (in two and three dimensions),
the stress threshold function $f_{\rm eigen}$
depends only on the {\em deviatoric} part of the stress
and can be written in the form:
\be
\label{feigen_geigen_incomp}
f_{\rm eigen}(\Sigma)=g_{\rm eigen}[\dev(\Sigma)]
\ee

As schematized on Fig.~\ref{quadrupoles_nombreux},
let us now discuss how the $T1$-threshold discussed above
impacts the plasticity threshold and the plastic flow.

\subsection{Plasticity threshold}
\label{plasticity_threshold}

\begin{figure}[ht!]
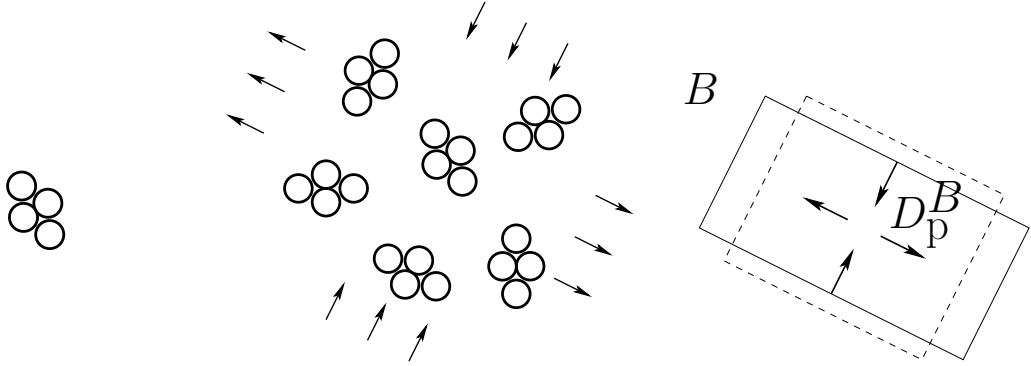

\begin{center}
\resizebox{0.25\columnwidth}{!}{\input{quadrupole_unique.pstex_t}}
\resizebox{0.38\columnwidth}{!}{\input{quadrupoles_nombreux.pstex_t}}
\resizebox{0.3\columnwidth}{!}{\input{quadrupole_to_continuum.pstex_t}}
\end{center}
\caption{From $T1$ process to plastic flow.
Left: the stress needed for four bubbles
to undergo a $T1$ process is illustrated
on Fig.~\protect{\ref{t1_df_dx}}.
Centre: when a chunk of material contains
a number of quadruplets of bubbles
with various orientations,
it may be postulated that the plasticity threshold
corresponds to the most favourably oriented quadruplet
(this assumption discards the effects
of non-homogeneous deformations,
and is discussed in the main text).
Right: the rate at which $T1$ processes occur
in the chunk of material when it is subjected
to elongation $\Finger$ is expressed
in the form of a plastic flow $\epspp$,
where the material configuration at time $t$ (dashed line)
is mapped onto configuration at time $t+{\rm d}t$ (full line)
through infinitesimal deformation $\epspp{\rm d}t$.%
}
\label{quadrupoles_nombreux}
\end{figure}

Instead of a single $T1$-capable quadruplet
(Fig.~\ref{quadrupoles_nombreux}, left), 
let us now consider a mesoscopic element of material,
containing quadruplets with many different orientations
(Fig.~\ref{quadrupoles_nombreux}, centre).

If we neglect the effects of disorder
(see next paragraph for a short discussion),
the plastic threshold for this chunk of material
is reached when the most favourably oriented quadruplet
reaches its own threshold:
\be
\label{f_sigma_average_alpha_comp}
f(\si)=\max_{\alpha}
f_{\rm eigen}(\si_\alpha)\nonumber
\ee
where $\alpha$ runs over all orientations in space,
and where $\si_\alpha$ is the representation of tensor $\si$
in the axes of a quadruplet oriented according to $\alpha$.
Note that the most sensitive quadruplets are those oriented along the stress:
the most favourable orientation $\alpha$ 
coincides with that of the stress tensor
(possibly up to some permutation of the axes).

In the case of the particular, two-dimensional threshold
given by Eqns.~(\ref{SiST12d}) and~(\ref{seuil_Sigma_2d}),
we obtain:
\be
\label{seuil_sigma_eigen}
f(\si)=|\si_{(1)}-\si_{(2)}|-\Sis(p)
\ee
where $\si_{(1)}$ and $\si_{(2)}$ are the eigenvalues 
of the stress tensor and where
$p=-\si_{(1)}-\si_{(2)}$ is the pressure.

In the case of an incompressible material,
the plastic threshold depends
on the sole {\em deviatoric} part of the stress,
like in Eq.~(\ref{feigen_geigen_incomp}):
\bee
f(\si)&=&\max_{\alpha}g_{\rm eigen}[\dev(\si_\alpha)]\nonumber\\
&=&g[\dev(\si)]
\label{f_sigma_average_alpha}
\eee

\subsection{Disorder and plasticity threshold}
\label{disorder}

The local disorder of soft object positions and interactions
implies that the stress acting on a particular quadruplet
slightly differs from the ambient mesoscopic stress,
and the threshold of some of the quadruplets is lower than expected
for a particular orientation of the applied stress.

As a result, the threshold value $\Sis$
in Eq.~(\ref{seuil_sigma_eigen}) above
is in fact slightly {\em lower} than the microscopic value $\Sis^{T1}$
in Eq.~(\ref{seuil_Sigma_2d}).
Similarly, expression~(\ref{f_sigma_average_alpha_comp}) for $f(\si)$
is slightly overestimated:
$\si_\alpha$ should in fact be understood
as the $\alpha$-oriented representation
of a locally {\em disorder-enhanced} version
of the mesoscopic stress $\si$.

\subsection{Plastic deformation rate}

In a continuum model,
the rate at which topological events occur
is expressed as the plastic flow $\epspp$
and depends on the stress $\si$,
or equivalently on the material deformation
(tensor $\Finger$)
{\em via} its elasticity (see Eq.~\ref{sdWdFinger}).
The plastic flow therefore has the following form:
\be
\label{eq_elast_plast}
\epspp(\si)=h(\Finger),
\ee
where $h$ is a tensorial function.
Since the plastic flow results from $T1$ events,
whose principal axes almost%
\footnote{up to the effect of disorder,
see paragraph~\ref{disorder}.}
coincide with those of the stress
(see paragraphs~\ref{T1_threshold}
and~\ref{plasticity_threshold}),
function $h$ is an isotropic function of $\Finger$:
it is such that the principal axes of $\epspp$
also coincide with those of $\Finger$.
Thus, in the same axes as those of $\Finger$:
\be
\label{epspp_delta_i}
\epspp=
\begin{pmatrix}
\delta_1&0&0\\
0&\delta_2&0\\
0&0&\delta_3
\end{pmatrix}
\ee
Equivalently, since powers of tensor $\Finger$
constitute a basis for symmetric tensors
that have the same principal axes,
$\epspp$ can be decomposed
for instance in the following way:
\be
\label{epspp_decomp_Finger_sans_dev}
\epspp=\bar{b}_0\,\unittensor
+\bar{b}_1\,\Finger
+\bar{b}_2\,\Finger^2,
\ee
where scalar functions $\bar{b}_0(\Finger)$, 
$\bar{b}_1(\Finger)$ and $\bar{b}_2(\Finger)$
depend solely on the scalar invariants of tensor $\Finger$.
Let us emphasize the fact that in general,
all three functions should be non-zero
above the plasticity threshold.
Restricting $\epspp$ to be proportional to $\Finger$, for instance,
would mean that all three directions would flow indepedently of each other.
This would be a very specific choice,
certainly not relevant for most materials.
It would be the plastic equivalent to taking
a Poisson ratio $\nu=0$ in linear elasticity,
a property valid for only a restricted class of materials.

Function $h$ (or equivalently {\em eigenvalues} $\delta_i$
in Eq.~\ref{epspp_delta_i}
or coefficients $\bar{b}_0$, $\bar{b}_1$ and $\bar{b}_2$
in Eq.~\ref{epspp_decomp_Finger_sans_dev}),
must also obey some constraints
based on physical grounds,
which are discussed in the next few paragraphs.

\subsection{Incompressible plastic flow rate}
\label{incompressible_plastic_flow_rate}

\begin{figure}[ht!]
\begin{center}
\resizebox{0.7\columnwidth}{!}{\input{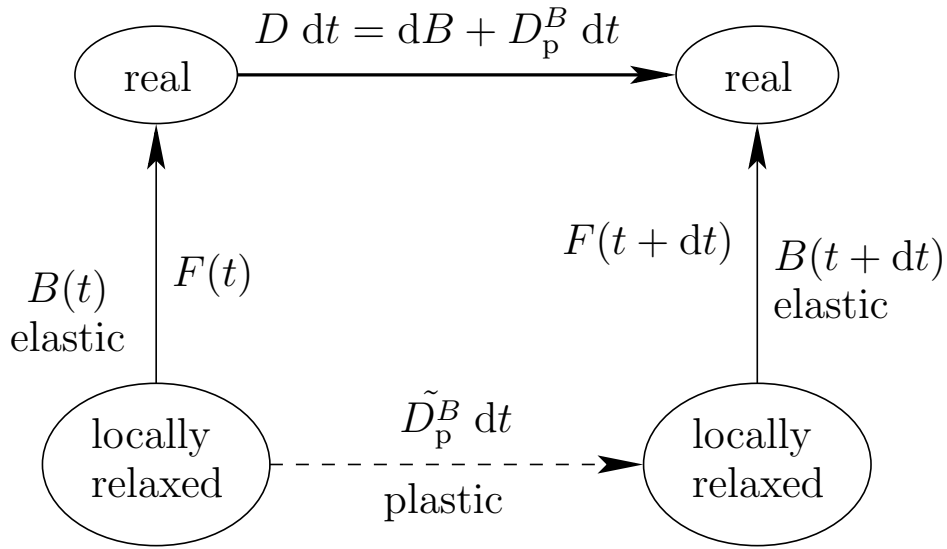}}
\end{center}
\caption{Incompressibility and plasticity.
The incompressible evolution $\epsppt$
of the relaxed conformation implies
the usual property $\trace(\epsppt)=0$.
Despite this, the corresponding flow $\epspp$
in real space is not traceless in general
($\trace(\epspp)\neq 0$): it is related to $\epsppt$
through a {\em finite} transformation $\Finger$,
see Eq.~(\protect{\ref{epsppt}}) for details.%
}
\label{deformation_elasto_plastique}
\end{figure}

Due to plasticity, the deformation between two real configurations
(see Figure~\ref{deformation_elasto_plastique})
does not reflect solely the increment 
$\frac{{\rm d}\Finger}{\dt}\dt$
in stored elastic deformation,
but also the irreversible drift $\epsppt\dt$
of the relaxed local configuration,
where the drift rate $\epsppt$ 
is related to the plastic term $\epspp$
defined by Eq.~(\ref{eq_elast_plast})
through covariant transport {\it via} tensor $F$:
\be
\label{epsppt}
\epsppt=\inv{\FF}\cdot\epspp\cdot\inv{\FF}
\ee

In the present work, we do not consider
all possible plastic deformation modes.
Among the situations depicted 
on Fig.~\ref{permeation_gas_diffusion_dilation},
conformations $(0)$, $(1)$ and $(4)$
are equilibrium situations
(at least on time scales where the dispersed phase does not diffuse
and where the continuous phase does not permeate).
Hence, modes $(0)\rightarrow(1)$ 
and $(0)\rightarrow(4)$ correspond
to plastic evolutions of the material.
In the present work, 
as stated in Section~\ref{evolution_modes},
such evolutions that involve mutual diffusion
between both phases are not addressed.

The plastic evolutions of interest for us
correspond to stress-induced evolutions,
where the relaxed conformation
is always characterized by the same amount
of material per bubble or droplet (inside and around it).
Among the equilibrium conformations $(0)$, $(1)$ and $(4)$,
only conformation $(0)$ is thus eligible.

We are thus interested in elastic deformations
(combinations of modes $(0)\rightarrow(2)$ 
and $(0)\rightarrow(3)$)
followed by plastic rearrangements
that bring the local conformation
back to (or at least {\em towards}) situation $(0)$.
In other words, the local relaxed conformation
is always locally similar to situation $(0)$,
even though some bubbles or droplets
may have swapped positions.

In particular, the evolution of the local relaxed state
must not be accompanied by any change in volume.
This can be expressed through a condition 
on the drift rate $\epsppt$:
\be
\trace(\epsppt)=0
\ee
Using Eq.~(\ref{epsppt}),
this can be expressed in terms of $\epspp$:
\bee
\label{incompressibilite_plastique}
0&=&\trace(\inv{\Finger}\cdot\epspp)\\
&=&\frac{\delta_1}{\beta_1}
+\frac{\delta_2}{\beta_2}
+\frac{\delta_3}{\beta_3}
\nonumber
\eee
Expression~(\ref{incompressibilite_plastique})
thus expresses incompressibility
for the plastic flow rate $\epspp$.

Note that in order to satisfy
the incompressibility condition 
given by Eq.~(\ref{incompressibilite_plastique}),
the plastic flow 
given by Eq.~(\ref{epspp_decomp_Finger_sans_dev})
can be written in the form:
\be
\label{epspp_decomp_Finger}
\epspp=b_1\,\Finger\cdot\dev(\Finger)
+b_2\,\Finger\cdot\dev(\Finger^2)
\ee
Scalar functions $b_1$ and $b_2$, 
like $\bar{b}_0$, $\bar{b}_1$ and $\bar{b}_2$
in Eq.~(\ref{epspp_decomp_Finger_sans_dev}),
depend solely on the invariants of tensor $\Finger$.

In this case, the coefficients 
of Eq.~(\ref{epspp_decomp_Finger_sans_dev})
can be obtained through the relations:
\bee
\bar{b}_0&=&b_2\\
\bar{b}_1&=&-\frac{b_2}{2}[\trace(\Finger)]^2
+\frac{b_2}{6}\trace(\Finger^2)
-\frac{b_1}{3}\trace(\Finger)\\
\bar{b}_2&=&b_1+b_2\;\trace(\Finger)
\eee
and are related to one another through:
\be
\frac{\bar{b}_0}{6}
\left([\trace(\Finger)]^2-\trace(\Finger^2)\right)
+\bar{b}_1
+\frac{\bar{b}_2}{3}\trace(\Finger)
=0
\ee

\subsection{Elastic versus plastic incompressibility}

At this point, it may be useful to precisely delineate
two different types of incompressibility,
which can be discussed 
using Fig.~\ref{permeation_gas_diffusion_dilation}.

As stated in paragraph~\ref{incompressible_plastic_flow_rate} above,
in the present work we are only interested
in {\em locally relaxed states} of type $(0)$,
and the plastic flows we consider
correspond to relaxed conformations that evolve
among such states of type $(0)$,
and the density of the locally relaxed state is conserved:
the material is {\em plastically incompressible}.

By contrast, as mentioned in paragraph~\ref{evolution_modes},
the {\em actual} state of the system (with local stresses)
may have a local density that differs from that of the relaxed state,
due to evolutions of type $(0)\rightarrow(2)$.
In other words, the material is assumed to be compressible,
that is: {\em elastically compressible}.\footnote{Note 
that this elastic compressibility
can always be defined on short time scales,
even if the threshold stress for plasticity
is arbitrarily small.}

It follows that with our assumptions,
the material may locally change volume
(and the density $\rho$ then departs from its initial value $\rho_0$)
when the flow is not divergence-free
($\nabla\cdot\vec{v}=\trace\epsp \neq 0$).
Nevertheless, because the plastic evolution is assumed incompressible,
the material returns to its initial density $\rho_0$
as soon as the local stress vanishes.

For some systems, elastic compressibility is negligible
and the material can be considered elastically incompressible,
as explained in paragraph~\ref{compressibility}.
The material then locally never changes volume,
and the flow is then diver\-gence-free
($\nabla\cdot\vec{v}=\trace\epsp=0$).

\subsection{Thermodynamic constraints}

The total work developed by the stress in the material
is given by $\trace(\si\cdot D)$.
In the present system, the internal energy $\energy$
is purely elastic: $\energy=\W$.
The first principle of thermodynamics
can thus be written as
\be
\rho\;{\rm d}\W=\trace(\si\cdot D)\,{\rm d}t+\delta Q,
\ee
where $\delta Q$ is the heat uptake per unit volume.

Besides, the dissipated power $\pdissip$ per unit volume
results from viscosity, especially during the relaxation
of individual $T1$ processes.
It constitutes the only source of entropy in the system,
which is non-negative according to the second principle:
\be
\pdissip=T\,{\rm d}\screated\geq 0
\ee
It also constitutes the only source of heat in the system.
If we assume that the system remains at a constant temperature,
this condition can be written as:
\be
\pdissip\,{\rm d}t+\delta Q=0
\ee

The Clausius-Duhem inequality
is readily derived from the above equations:
\be
\label{ClDu}
\trace(\si\cdot D)-\rho\;\frac{{\rm d}\W}{{\rm d}t}
\geq 0
\ee
Let us express this inequality in a different way.

From Eq.~(\ref{sidWdFinger}), we get:
\be
\label{trsiD}
\trace(\si\cdot D)=2\rho\,\trace\left(
\Finger\cdot\frac{{\rm d}\W}{{\rm d}\Finger}\cdot D
\right)
\ee
We also have:
\be
\label{dWdt}
\frac{{\rm d}\W}{{\rm d}t}
=\trace\left(\frac{{\rm d}\W}{{\rm d}\Finger}
\cdot\frac{{\rm d}\Finger}{{\rm d}t}\right)
\ee
where ${\rm d}\Finger/{\rm d}t$ is given by
Eq.~(\ref{dFingerplast}).
Hence, from Eqs.~(\ref{ClDu}), (\ref{trsiD}) and~(\ref{dWdt}),
we obtain the constraint on the plastic flow $\epspp$
that corresponds to the Clausius-Duhem inequality:
\be
\label{dissipation_plastique_positive}
\pdissip=2\rho\;\trace\left(
\epspp\cdot\frac{{\rm d}\W}{{\rm d}\Finger}
\right) \geq 0
\ee

\subsection{Plastic deformation rate: a summary}

In the last few paragraphs, we showed
that the plastic deformation rate
must obey constraints that express the fact that:
\begin{enumerate}
\item the plastic flow depends on the stress
or equivalently on the stored deformation
(see Eqs.~\ref{eq_elast_plast} 
or~\ref{epspp_decomp_Finger_sans_dev});
\item if applicable, the plastic flow is incompressible
(see Eqs.~\ref{incompressibilite_plastique}
or~\ref{epspp_decomp_Finger});
\item the associated dissipation is positive
(see Eq.~\ref{dissipation_plastique_positive}).
\end{enumerate}

\section{Complete, continuous model}
\label{complete_continuous_model}

In this brief section, let us discuss
how the constitutive equation derived in the present work
can be inserted into a set of equations
and provide a complete, continuous model.

Keeping in mind the considerations 
of paragraph~\ref{two_phase_fluid},
let us now close the evolution equation
of the system (Eqs.~\ref{dFingerplast}
or~\ref{dedtplast})
\be
\frac{{\rm d}\Finger}{{\rm d}t}
-\gradv \cdot \Finger
-\Finger \cdot \transp{\gradv}
=-2\,\epspp
\ee
One needs the elastic law
(Eq.~\ref{elasticLawIBB2} or~\ref{sidWdFinger})
\be
\label{stress_elastique_section_dissipation}
\si=2\,\rho\,\frac{{\rm d}\W}{{\rm d}\Finger}\cdot\Finger
\ee
and the usual force balance equation
\be
\label{force_balance_equation}
\nabla\cdot\si+\rho\,\vec{f}
=\rho\frac{{\rm d}\vec{v}}{{\rm d}t},
\ee
where $\vec{f}$ represents external forces per unit mass.
The evolution of the density $\rho$
obeys the usual mass conservation equation
\be
\label{density_evolution}
\frac{\partial\rho}{\partial t}
+\nabla\cdot(\rho\;\vec{v})
=\frac{{\rm d}\rho}{\dt}
+\rho\,\trace\epsp
=0
\ee
We provide later
(see Eq.~\ref{density_evolution_with_permeation})
a version of this equation 
that includes liquid permeation.

Let us now mention the special case
of an initially non-dilated material,
and discuss dissipation at weak applied stresses.

\subsection{Initially non-dilated material}

Let us assume that the material 
initially has a uniform density
{\em i.e.,} with stored deformation
\be
\rho(t_0,\vec{r})=\rho_0
\ee
and that it verifies everywhere
\be
\det\Finger(t_0,\vec{r})=1,
\ee

In such a case, our assumptions stated in paragraph~\ref{evolution_modes}
imply that the material density
is related to the determinant of tensor $\Finger$
in a very simply way.
Indeed, the derivation 
shown in Appendix~\ref{density_and_stored_deformation}
implies that at any later time, throughout the sample:
\be
\label{rho_inv_sqrt_Finger}
\rho
=\frac{\rho_0}
{\sqrt{\det\Finger}}
\ee

With such assumptions,
Eq.~(\ref{rho_inv_sqrt_Finger})
can therefore be used to replace $\rho$
within Eq.~(\ref{force_balance_equation}),
and Eq.~(\ref{density_evolution})
is then not useful any more.

\subsection{Viscous losses under weak stresses}

The Bingham model addressed in the present work
reduces to a simple elastic system
when subjected to weak stresses
(below the yield stress).
As a result, vibrations may be present
in the material, at arbitrary high frequencies.
Such vibrations may be undesirable.
Not only do they make numerical simulations
of the above system of equations more complicated,
if not impossible,
but they do not faithfully reflect
the damping observed in real materials.
This problem does not arise under large stresses,
as the plastic flow rate introduces dissipation.

One of the simplest ways to introduce 
some dissipation at weak stresses
is to add a viscous term to the stress.
Eq.~(\ref{stress_elastique_section_dissipation})
thus becomes:
\be
\si=2\,\rho\,\frac{{\rm d}\W}{{\rm d}\Finger}\cdot\Finger
+V(\epsp,\;\Finger)
\ee
where the viscous term $V$ is a symmetric tensor
which depends linearly\footnote{We assume that all
non-linearities in $\epsp$ are included in the dependence
on the deformation $\Finger$,
which itself results from the applied deformation rate $\epsp$.} 
on the applied deformation rate $\epsp$.
As the material is locally anisotropic 
due to the stored deformation,
$V$ additionally depends on $\Finger$.

The general form of function $V(\epsp,\;\Finger)$
is a sum of terms whose principal axes
are those of tensor $\Finger$:
\be
l_1(\Finger)\,\trace(l_2(\Finger)\cdot\epsp)
\ee
and of terms that depend tensorially on $\epsp$:
\be
m_1(\Finger)\cdot\epsp\cdot m_2(\Finger)
+m_2(\Finger)\cdot\epsp\cdot m_1(\Finger)
\ee
Here, all functions $l_i$ and $m_i$
are isotropic scalar functions of tensor $\Finger$.

If we neglect any effect of the material deformation,
we may simply add a linear viscous term,
and Eq.~(\ref{stress_elastique_section_dissipation})
then becomes:
\be
\si=2\,\rho\,\frac{{\rm d}\W}{{\rm d}\Finger}\cdot\Finger
+\eta\;\epsp
\ee
%

\section{Homogeneous constant shear flow: one example}
\label{homogeneous_constant_shear_flow}

Let us now use the constitutive Eq.~(\ref{dFingerplast})
to obtain the evolution of the system
in a very common type of flow: a shear flow
at a constant shear rate $\gd$
starting at time $t=0$. 
For non-homogeneous flows,
the constitutive equation
must be combined with the classical mechanics equations
for continuum media,
as mentioned in Section~\ref{complete_continuous_model}.
Here, for simplicity, we assume
that the material is homogeneous and incompressible
(uniform density $\rho_0$ at all times)
and that the flow remains homogeneous:
no shear-banding, etc.

\begin{figure}[h!]
\begin{center}
\resizebox{0.8\columnwidth}{!}{%
  \input{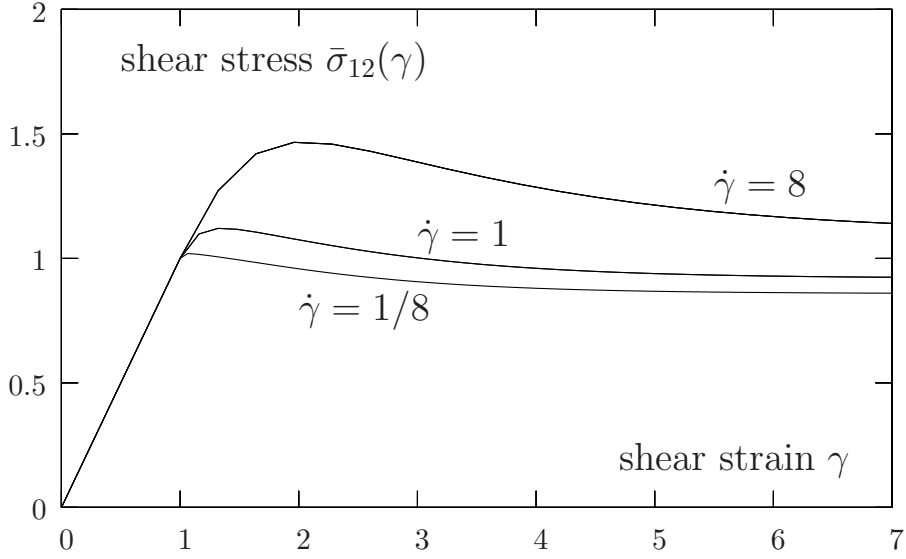}
}
\end{center}
\caption{Shear stress $\sud$ in the course of a shear experiment,
as a function of the shear deformation $\gamma$,
for three different shear rate values $\gd$.
The elasticity and plasticity terms are chosen as described 
in Section~\ref{chosen_elasticity_threshold_and_plastic_flow}.
The response increases during the initial period of time,
when the material deforms purely elastically
(as expected, the response is then independent of the shear rate).
The increase then levels off
once the plasticity threshold has been reached,
and saturates at a stationary value.
For a plot of the final, 
stationary value on the applied shear rate,
see Fig.~\ref{cisaillement_permanent_s12_k1_17_k2_67}.}
\label{cisaillement_evolution_s12_k1_17_k2_67}
\end{figure}

\begin{figure}[h!]
\begin{center}
\resizebox{0.8\columnwidth}{!}{%
  \input{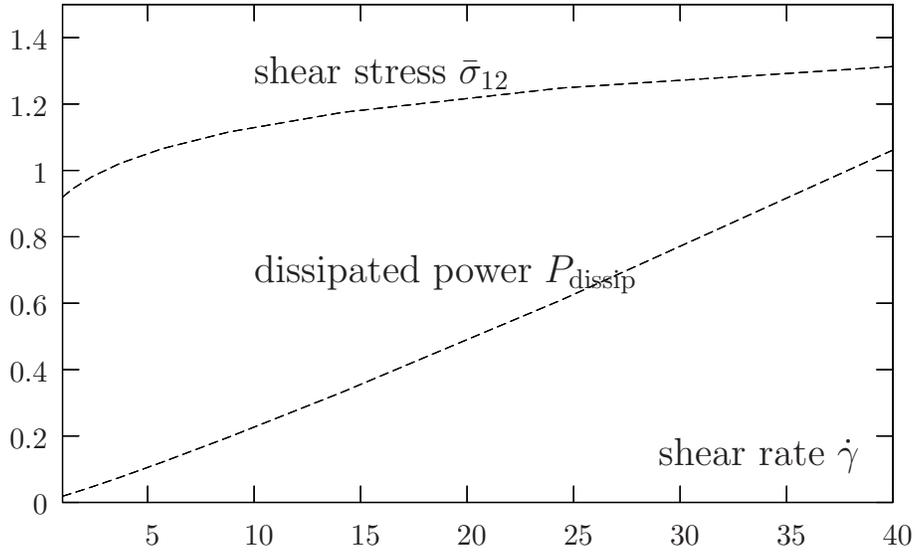}
}
\end{center}
\caption{Shear stress $\s12$ 
and dissipation power $\pdissip$ per unit volume
in the stationary regime,
as a function of the shear rate $\gd$.
The elasticity and plasticity terms are chosen as described 
in Section~\ref{chosen_elasticity_threshold_and_plastic_flow}.}
\label{cisaillement_permanent_s12_k1_17_k2_67}
\end{figure}

A typical result from Eq.~(\ref{dFingerplast}),
coupled with an elastic law 
(here incompressible, see Eq.~\ref{s_Finger})
is the shear stress as a function 
of the shear deformation $\gamma$
since time $t=0$ when shear started.
An example of such a mechanical response
(with parameters chosen as described 
in Section~\ref{chosen_elasticity_threshold_and_plastic_flow})
is provided on Fig.~\ref{cisaillement_evolution_s12_k1_17_k2_67}.
The equations of our model provide:
\begin{enumerate}
\item the material response in its elastic state
(first part of the curves on Fig.~\ref{cisaillement_evolution_s12_k1_17_k2_67});
\item the threshold that marks the onset of plasticity
(point where curves split apart depending on the shear rate);
\item the transient response that results from plasticity
(second part of the curves);
\item the stationary response,
as a function of the applied shear rate $\gd$
(see Fig.~\ref{cisaillement_permanent_s12_k1_17_k2_67}).
\end{enumerate}

\begin{figure}[h!]
\begin{center}
\resizebox{0.8\columnwidth}{!}{%
  \input{cisaillement_evolution_dissipation_k1_17_k2_67.pstex_t}
}
\end{center}
\caption{Dissipation $\pdissip$ per unit volume 
and injected power $\sud\;\gd$
in the course of a shear experiment,
as a function of the shear deformation $\gamma$.
The applied shear rate is $\gd=8$.
The elasticity and plasticity terms are chosen as described 
in Section~\ref{chosen_elasticity_threshold_and_plastic_flow}.
Note that there is no dissipation ($\pdissip=0$)
in the elastic regime, prior to the onset of plasticity.
Part of the injected work is not dissipated
(non-zero surface area between both curves)
and corresponds to the elastic energy
stored in the material.}
\label{cisaillement_evolution_dissipation_k1_17_k2_67}
\end{figure}

\subsection{Method}

\begin{figure}[h!]
\begin{center}
\resizebox{0.4\columnwidth}{!}{%
  \input{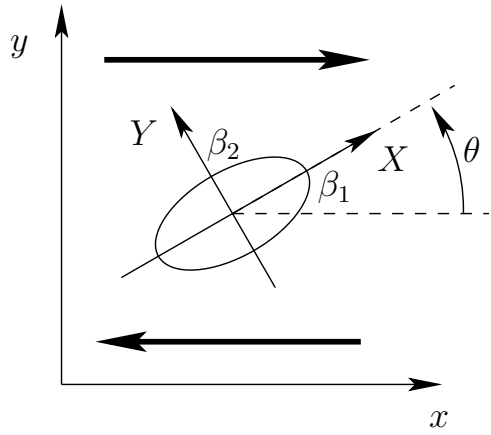}
}
\end{center}
\caption{The homogeneous shear flow imposed
in basis $xyz$ attached to the laboratory
($z$ being the direction of vorticity)
causes the principal axes of tensor $\Finger$
to tilt (basis $XYz$) with respect to the shear.
The lengths of the ellipse axes represent 
the magnitudes of eigenvalues $\beta_1$ and $\beta_2$.}
\label{xyXYz}
\end{figure}

Let us recall the evolution equation~(\ref{dFingerplast})
for the material deformation (tensor $\Finger$):
\be
\label{dFingerplast2}
\frac{{\rm d}\Finger}{{\rm d}t}
-\gradv \cdot \Finger
-\Finger \cdot \transp{\gradv}
=-2\,\epspp
\ee
This equation must be understood as written
in a basis attached to the (fixed) laboratory frame.
Let $x$ be the axis of velocity,
$y$ the axis of the velocity gradient
and $z$ the vorticity axis.
The symmetry of the shear flow implies that
the corresponding material deformation $\Finger$
has two principal axes within the $xy$ plane
(in directions $X$ and $Y$ yet to be determined,
see Fig.~\ref{xyXYz})
and one principal axis along $z$.
Note that $\epspp$ and $\s$ also has
the same principal axes $X$, $Y$ and $z$ as $\Finger$
(see Eqs.~\ref{sidWdFinger} and~\ref{epspp_decomp_Finger}).

As shown in Appendix~\ref{derivation},
Eq.~(\ref{dFingerplast2})
provides a system of differential equations for $\beta_1$, $\beta_2$ and for angle $\theta$
between axes $x$ and $y$ (see Fig.~\ref{xyXYz}):
\bee
\label{cos2theta}
u&=&\cos(2\theta)\\
\label{dbeta1dt}
\frac{{\rm d}\beta_1}{{\rm d}t}&=&\gd\beta_1\sqrt{1-u^2}-2\delta_1(\beta_1,\beta_2)\\
\label{dbeta2dt}
\frac{{\rm d}\beta_2}{{\rm d}t}&=&-\gd\beta_2\sqrt{1-u^2}-2\delta_2(\beta_1,\beta_2)\\
\label{dcos2thetadt}
\frac{{\rm d}u}{{\rm d}t}&=&\gd\sqrt{1-u^2}
\left\{1-u\frac{\beta_1+\beta_2}{\beta_1-\beta_2}\right\}\\
\label{beta3equals1overbeta1beta2}
\beta_3&=&\frac{1}{\beta_1 \beta_2}
\eee
where Eq.~(\ref{beta3equals1overbeta1beta2})
results from the assumed material incompresibility.

The above equations provide the evolution
of the material deformation (tensor $\Finger$)
from the initial situation at rest\footnote{See
Appendix~\ref{elasticshearflow} for the apparent singularity
at $t=0$ when $\beta_1=\beta_2$
and for the state of the system in the elastic regime.}
to the onset of plasticity, 
the plastic transient
and the final, stationary state 
(see Section~\ref{stationaryshearflow} of Appendix).

Once the evolution of tensor $\Finger$ is known,
the stress is obtained through Eq.~(\ref{sdWdFinger}).

As for dissipation, given by 
Eq.~(\ref{dissipation_plastique_positive}),
it involves both the gradient of the elastic energy
(see Eq.~\ref{dWdFinger})
and the plastic deformation rate
given by Eq.~(\ref{epspp_decomp_Finger}).

The dissipation per unit volume can now be expressed as:
\bee
\pdissip&=&2\rho\,\trace\left(
\epspp\cdot\frac{{\rm d}\W}{{\rm d}\Finger}
\right)\\
&=&3a_1\bar{b}_0\nonumber\\
&&+(a_2\bar{b}_0+a_1\bar{b}_1)\;\trace(\Finger)\nonumber\\
&&+(a_1\bar{b}_2+a_2\bar{b}_1)\;\trace(\Finger^2)\nonumber\\
&&+a_2\bar{b}_2\;\trace(\Finger^3)
\eee
When $\epspp$ is known explicitely,
it can be calculated more directly:
\bee
\pdissip&=&
(a_1+a_2\beta_1)\;\delta_1\nonumber\\
&+&(a_1+a_2\beta_2)\;\delta_2\nonumber\\
&+&(a_1+a_2\beta_3)\;\delta_3
\eee

Figure~\ref{cisaillement_evolution_dissipation_k1_17_k2_67}
displays both the injected power, $\s12\;\gd$,
and the dissipated power, $\pdissip$,
in the course of a shear experiment.
Notice that as long as the material 
remains in the elastic regime,
no dissipation occurs.
The injected power is being stored entirely
as elastic energy.
As dissipation starts, an overshoot of dissipated power is observed.
Asymptotically, both quantities converge
towards the same value
while the elastic deformation and energy of the material
reach their stationary values.

\subsection{Chosen elasticity, threshold and plastic flow}
\label{chosen_elasticity_threshold_and_plastic_flow}

Let us choose a very common (Mooney-Rivlin) type 
of incompressible elasticity,
which has been shown to suitably approximate
the non-linear elasticity of 
foams~\cite{hoehler_cohen_addad_review_2005,hoehler_cohenaddad_rabiausse_2004}.
The corresponding elastic energy~\cite{bertram_2005} 
can be expressed as:
\be
\rho_0\,\W=\frac{\mra}{2}({\rm I}_\Finger-3)
+\frac{\mrb}{2}({\rm II}_\Finger-3)
\ee
where
\bee
{\rm I}_\Finger&=&\trace(\Finger)\\
{\rm II}_\Finger&=&\frac12[\trace^2(\Finger)-\trace(\Finger^2)]
=\trace(\Finger^{-1})
\eee
Correspondingly, the coefficients
of Eq.~(\ref{s_Finger})
can be expressed as:
\bee
\label{a1_a2_mooney_rivlin}
a_1&=&\mra+\mrb\,{\rm I}_\Finger\\
a_2&=&-\mrb
\eee
As in refs.~\cite{hoehler_cohen_addad_review_2005,hoehler_cohenaddad_rabiausse_2004},
we choose the values of $\mra$ and $\mrb$
in terms of the shear modulus $\G$ as:
\bee
\mra&=&\frac17\G\\
\mrb&=&\frac67\G
\eee
Note that in Section~\ref{impact_elasticity_plasticity},
we explore other values for $\mra$ and $\mrb$
(still in the framework of a Mooney-Rivlin elasticity).

For the plastic deformation rate $\epspp$,
as expressed by Eq.~(\ref{epspp_decomp_Finger})
so as to obey plastic incompressibility,
we choose the following coefficient values:
\bee
b_1&=&(\trace\Finger-4)\;\theta(\trace\Finger-4)\\
b_2&=&0
\eee
where $\theta(x)=1$ when $x\geq0$ and $\theta(x)=0$ otherwise.
Note that in Section~\ref{impact_elasticity_plasticity},
we explore the case $b_2=b_1$ for comparison.

This choice implies, in particular,
that the plasticity threshold
corresponds to the following condition:
\be
\label{threshold_function}
\trace(\Finger-4)=0
\ee

We now discuss qualitatively the evolution 
of tensor $\Finger$.
We then show the effects of the material elasticity
and discuss the corresponding rheological response.

\subsection{Three-dimensional evolution of the material stored deformation}

\begin{figure}[h!]
\begin{center}
\resizebox{1.0\columnwidth}{!}{%
  \input{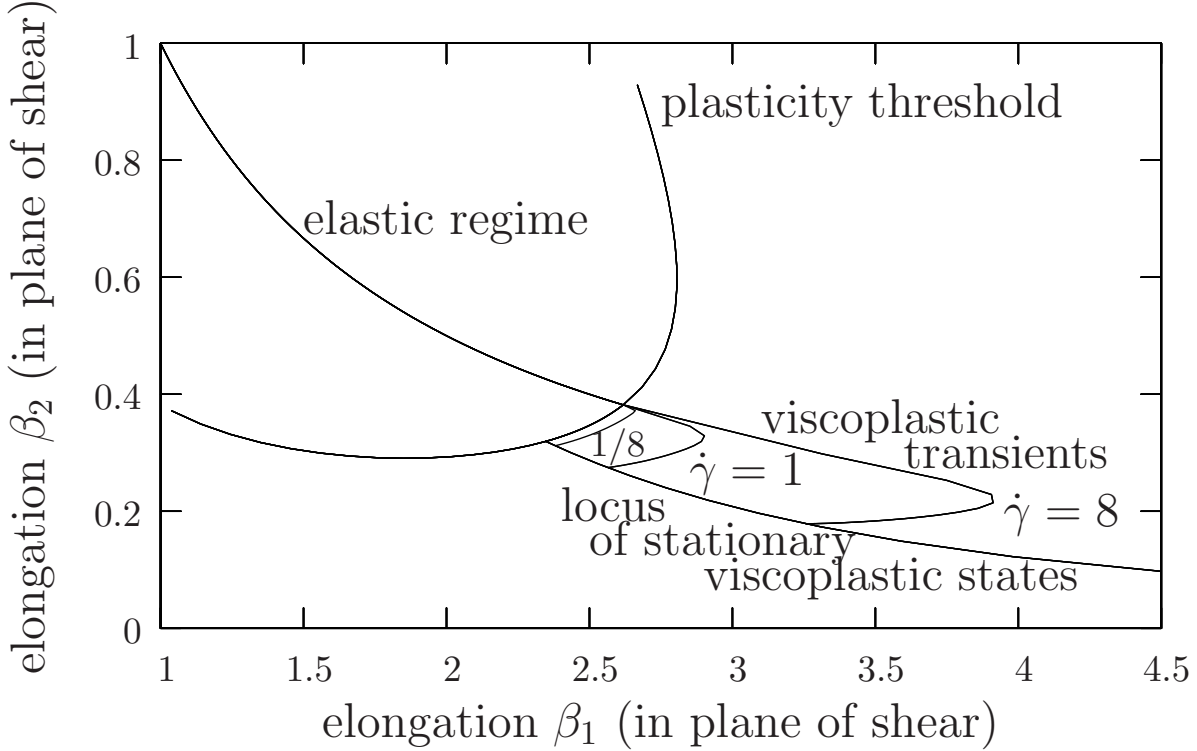}
}
\end{center}
\caption{Form of the stored elastic deformation
in the course of an experiment
and in the stationary regime,
for three different values
of the shear rate $\gd$.
The axes are the first two eigenvalues,
$\beta_1$ and $\beta_2$, of tensor $\Finger$.}
\label{cisaillement_beta1_beta2_k1_17_k2_67}
\end{figure}

With the choices made in 
Paragraph~\ref{chosen_elasticity_threshold_and_plastic_flow}
for the material elasticity and plasticity,
the evolution of the system 
given by Eqs.~(\ref{cos2theta})--(\ref{beta3equals1overbeta1beta2})
is depicted on Fig.~(\ref{cisaillement_beta1_beta2_k1_17_k2_67}).
For three different values
of the shear rate $\gd$,
it represents the trajectory of the material
stored deformation as for $\beta_1$ and $\beta_2$,
successively in the elastic regime
and in the plastic regime.
The plasticity threshold, 
given by Eq.~(\ref{threshold_function})
is also represented,
as well as the locus of the stationary states.

In the elastic regime, the form of 
Eqs.~(\ref{B_form_xyz})--(\ref{Dp_form_xyz})
imply that $\beta_3$ remains equal
to its initial value, $\beta_3=1$,
{\em i.e.}, that the product $\beta_1\beta_2$
remains equal to unity.
This can indeed be checked from
Eqs.~(\ref{beta1el})--(\ref{beta2el}),
and it reflects the fact that as long as
no plastic events have occured in the material,
the absence of material deformation in direction $z$
(planar shear in the $x$-$y$ plane, 
see Eq.~\ref{gradv_form_xyz})
implies that the stored deformation
is not modified in direction $z$.

In other words, in the elastic regime, the fact 
that no plasticity occurs in direction $z$
({\em i.e.}, $\delta_3=0$) 
implies that in the same direction,
the stored deformation remains constant, {\em i.e.},
$\beta_3={\rm const}$.

Conversely, in the stationary regime,
$\beta_3$ remains constant (like most other quantities)
and this imposes that no plasticity occurs in direction $z$,
{\em i.e.}, $\delta_3=0$.

Meanwhile, in the transient regime,
the stored deformation $\beta_3$ in direction $z$
has evolved from its initial value
towards its new, stationary value, despite the absence 
of any velocity gradient in this direction.
This is made possible by the plastic events ($\delta_3\neq 0$)
which allow internal relaxation within the material
even in the absence of any flow in this direction.

\begin{figure}[h!]
\begin{center}
\resizebox{0.8\columnwidth}{!}{%
  \input{cisaillement_evolution_betai_k1_17_k2_67.pstex_t}
}
\end{center}
\caption{Eigenvalues of the stored elastic deformation
in the course of an experiment
as a function of the shear deformation $\gamma$.
The applied shear rate is $\gd=8$.}
\label{cisaillement_evolution_betai_k1_17_k2_67}
\end{figure}

\begin{figure}[h!]
\begin{center}
\resizebox{0.8\columnwidth}{!}{%
  \input{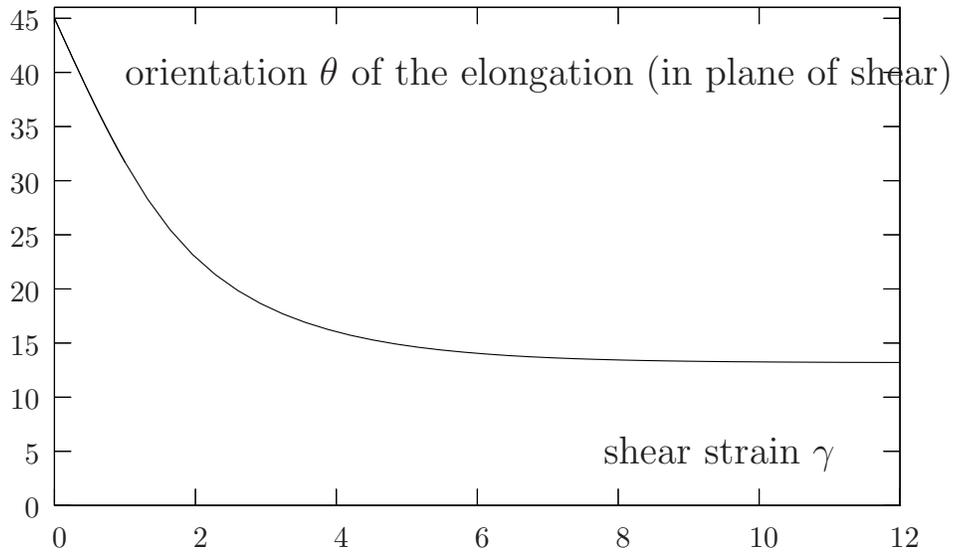}
}
\end{center}
\caption{Orientation of the stored elastic deformation
in the course of an experiment
as a function of the shear deformation $\gamma$.
The applied shear rate is $\gd=8$.}
\label{cisaillement_evolution_theta_k1_17_k2_67}
\end{figure}

\subsection{Shear thinning}

The stationary shear stress is represented on
Fig.~\ref{cisaillement_permanent_s12_k1_17_k2_67}
as a function of the shear rate $\gd$.
Above the yield stress,
the stress increases when $\gd$ is increased, as expected.
Notice that it increases in a subliminear way.
We believe that the main reason for this shear-thinning behaviour
could be the fact that the plastic flow rate we chose,
as expressed in termes $\Finger$,
is cubic rather than linear for large stored deformations:
\be
\epspp=(\trace\Finger-4)\;\theta(\trace\Finger-4)
\;\left[\Finger^2-\frac{\Finger}{3}\trace(\Finger)\right]
\ee
where, again, $\theta(x)=1$ when $x\geq0$ and $\theta(x)=0$ otherwise.

\subsection{Normal stress differences}

The stress tensor $\s$ is obtained from the Finger tensor $\Finger$
through Eq.~(\ref{s_Finger}).
In the case of a homogeneous shear flow (see above),
it can be expressed not only in the $XYz$ basis
(Eq.~\ref{s_Finger}) but also in the $xyz$ basis
associated with the flow:
\bee
\s&=&\begin{pmatrix}\suu&\sud&0\\
\sud&\sdd&0\\
0&0&\strois
\end{pmatrix}\nonumber\\
&=&\begin{pmatrix}c^2\su+s^2\sd
&cs(\su-\sd)&0
\\
cs(\su-\sd)&s^2\su+c^2\sd&0
\\ 0&0&\strois
\end{pmatrix}
\label{s_xyz}
\eee

where $s$ (resp., $c$) denotes the sine (resp., the cosine)
of angle $\theta$ between axes $x$ and $X$
(see Fig.~\ref{xyXYz}).

The first and second normal stress difference can be expressed as:
\bee
N_1&\equiv&\suu-\sdd\nonumber\\
&=&(\su-\sd)u\\
N_2&\equiv&\sdd-\strois\nonumber\\
&=&\frac{1-u}{2}(\su-\strois)+\frac{1+u}{2}(\sd-\strois)
\eee
with ${\s}_i$ given by Eq.~(\ref{s_Finger}):
\be
{\s}_i-{\s}_j=a_1\,(\beta_i-\beta_j)+a_2\,(\beta_i^2-\beta_j^2)
\ee
Hence, for instance:
\bee
N_1&=&u\left[a_1(\beta_1-\beta_2)+a_2(\beta_1^2-\beta_2^2)\right]\\
N_2&=&\frac{a_1}{2}\left[\beta_1+\beta_2
-\frac{2}{\beta_1\beta_2}+u(\beta_2-\beta_1)\right]\nonumber\\
&&+\frac{a_2}{2}\left[\beta_1^2+\beta_2^2
-\frac{2}{(\beta_1\beta_2)^2}
+u(\beta_2^2-\beta_1^2)\right]
\eee

\begin{figure}[h!]
\begin{center}
\resizebox{0.8\columnwidth}{!}{%
  \input{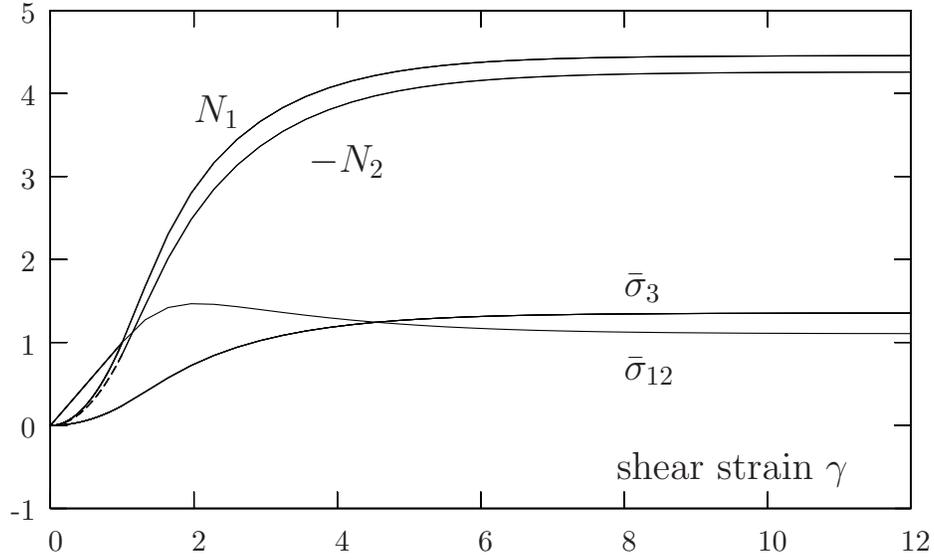}
}
\end{center}
\caption{Shear stress $\sud$, 
deviatoric stress $\strois$ in the vorticity direction
and normal stress differences $N_1$ and $N_2$
in the course of an experiment
as a function of the shear deformation $\gamma$,
for $\gd=8$.}
\label{cisaillement_evolution_contraintes_normales_k1_17_k2_67}
\end{figure}

\begin{figure}[h!]
\begin{center}
\resizebox{0.8\columnwidth}{!}{%
  \input{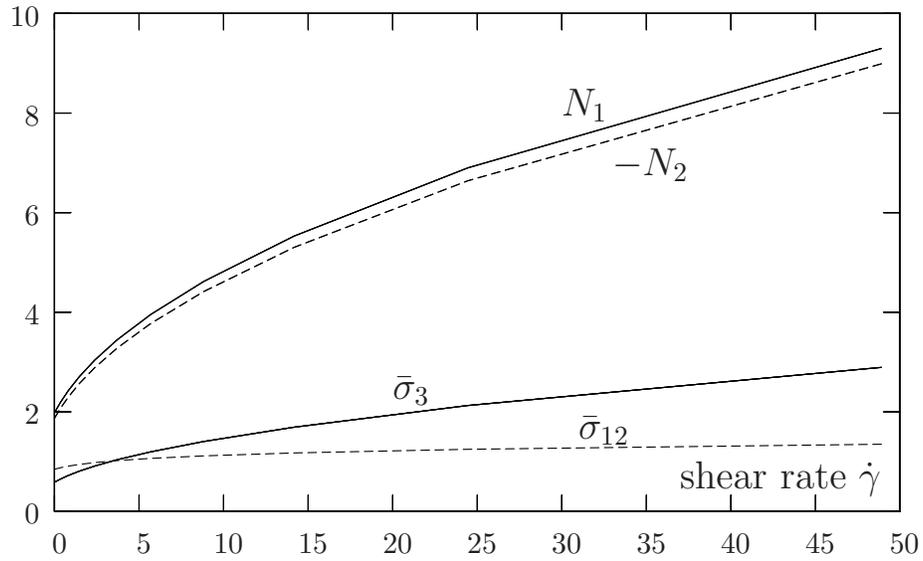}
}
\end{center}
\caption{Shear stress $\sud$, 
deviatoric stress $\strois$ in the vorticity direction
and normal stress differences $N_1$ and $N_2$
in the stationary regime,
as a function of the shear rate $\gd$.}
\label{cisaillement_permanent_contraintes_normales_k1_17_k2_67}
\end{figure}

In the case of Mooney-Rivlin,
coefficients $a_1$ and $a_2$ are given
by Eq.~(\ref{a1_a2_mooney_rivlin})
and the normal stress differences
can be expressed as:

\bee
N_1&=&u(\beta_1-\beta_2)
\left[\mra+\frac{\mrb}{\beta_1\beta_2}\right]
\label{N1_MR_shear}\\
N_2&=&\frac{\mra}{2}\left[\beta_1+\beta_2
-\frac{2}{\beta_1\beta_2}+u(\beta_2-\beta_1)\right]\nonumber\\
&&+\frac{\mrb}{2}\left[2\beta_1\beta_2-\frac{\beta_1+\beta_2}{\beta_1\beta_2}
+u\frac{\beta_2-\beta_1}{\beta_1\beta_2}\right]
\label{N2_MR_shear}
\eee

\subsection{Discussion of the stress response}

Let us now comment briefly
on the results presented 
on Fig.~\ref{cisaillement_evolution_contraintes_normales_k1_17_k2_67}.

Normal stress differences $N_1$ and $N_2$
increase gently and monotonically
to reach their stationary values.

The material is under traction
in the direction of vorticity ($\strois>0$),
as well as in the direction of $\beta_1$,
while it is under compression 
in the direction of $\beta_2$ (not shown).
This is consistent with the fact
that it is stretched in the direction of vorticity
($\beta_3>1$).

The salient feature of these results
is that the shear stress $\sud$
presents an overshoot during its transient,
a behaviour which is intrinsically unstable
in a homogeneous material,
and could trigger flow localization.
Such an overshoot for the shear stress has been observed 
in sheared foams~\cite{khan_overshoot}.

\section{Impact of elasticity and plasticity on the stress response}
\label{impact_elasticity_plasticity}

Our dynamical equation was formulated 
in terms of tensor $\Finger$ (see Eq.~\ref{dFingerplast2}).
From the resulting evolution of $\Finger$, 
when can then derive the stress evolution using 
the elastic law (Eq.~\ref{elasticLawIBB2}). 
In this section, we explore how the choice 
for the elasticity and the plasticity 
impact the stress response.

\subsection{Impact of elasticity}
\label{impact_of_elasticity}

Note that in this paragraph the deformation history 
will be the same in all cases, 
and only the stress in the material will differ. 
Just like in Section~\ref{homogeneous_constant_shear_flow},
we restrict ourselves to a Mooney-Rivlin type of elasticity,
which is complex enough for illustrative purposes,
but which has no specific properties with regards to our problem.

The elasticity of dry foams
has been shown~\cite{hoehler_cohen_addad_review_2005,hoehler_cohenaddad_rabiausse_2004}
to be well captured by such a Mooney-Rivlin elasticity,
using parameter values $k_1=1/7$ and $k_2=6/7$.
These values correspond to 
Figs.~\ref{cisaillement_evolution_s12_k1_17_k2_67}--%
\ref{cisaillement_permanent_contraintes_normales_k1_17_k2_67} 
where we took the modulus value $\G=1$.

\begin{figure*}
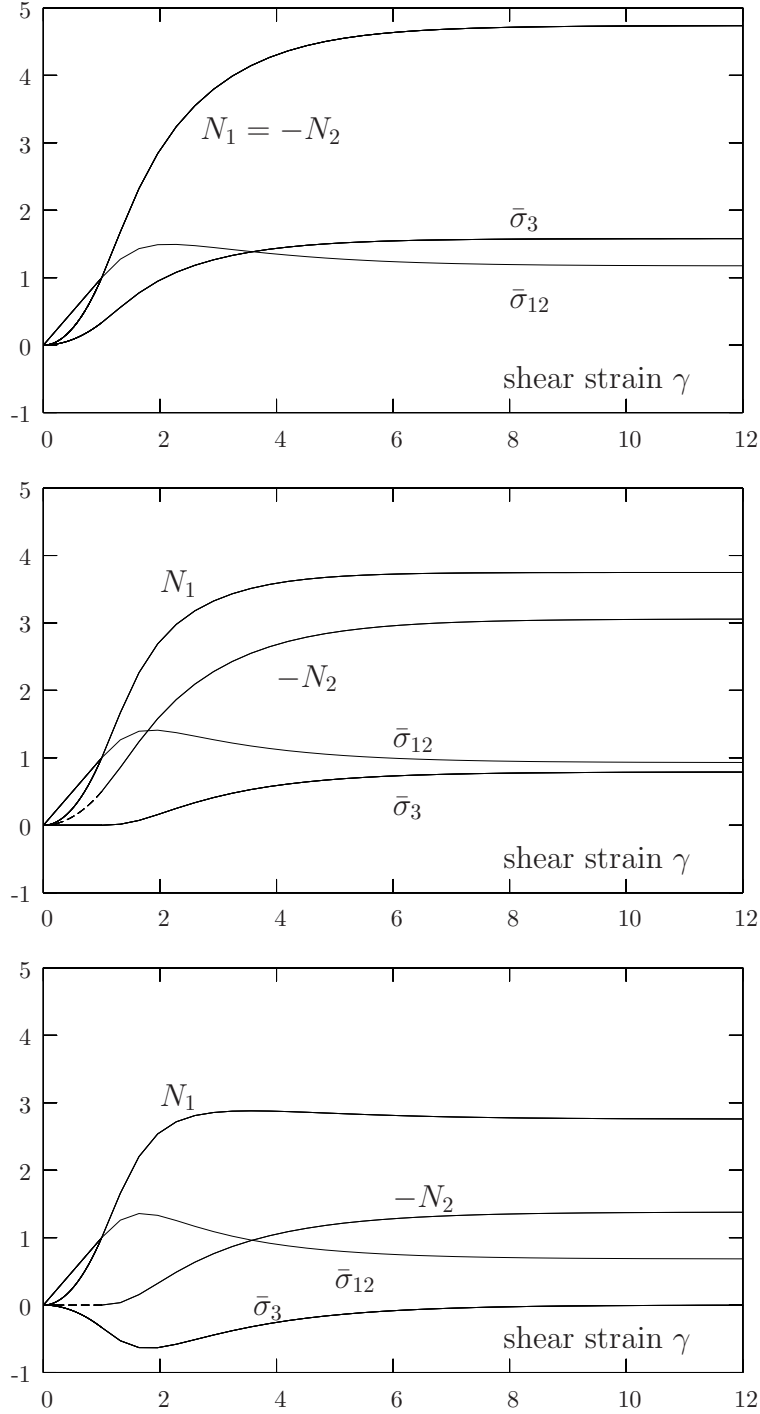

\begin{center}
\resizebox{0.65\columnwidth}{!}{%
  \input{cisaillement_evolution_contraintes_normales_k1_0_k2_1.pstex_t}
}
\resizebox{0.65\columnwidth}{!}{%
  \input{cisaillement_evolution_contraintes_normales_k1_12_k2_12.pstex_t}
}
\resizebox{0.65\columnwidth}{!}{%
  \input{cisaillement_evolution_contraintes_normales_k1_1_k2_0.pstex_t}
}
\end{center}
\caption{Impact of elasticity on the stress response.
The shear stress $\sud$, 
the deviatoric stress $\strois$ in the vorticity direction
and the normal stress differences $N_1$ and $N_2$
are plotted as a function of the shear deformation $\gamma$,
for a shear rate $\gd=8$.
Mooney-Rivlin elasticity is chosen with different parameter values.
Left: $k_1=0$ and $k_2=1$. 
Center: $k_1=1/2$ and $k_2=1/2$.
Right: $k_1=1$ and $k_2=0$.}
\label{cisaillement_evolution_contraintes_normales_k1_1_k2_0}
\end{figure*}

Exploring different values for $k_1$ and $k_2$ (still with $\G=1$),
we show in Fig.~\ref{cisaillement_evolution_contraintes_normales_k1_1_k2_0}
the time evolution of shear stress, stress in the vorticity direction
and normal stress differences.
The mechanical behaviour remains similar in all these cases.
In particular, the shear stress always presents an overshoot.
This overshoot is more pronounced 
for $\mra=1$ and $\mrb=0$.
In all cases, the normal stress differences
do not present any overshoot
and reach monotonically their stationary values.
As for the stress along the vorticity direction, 
It presents an undershoot when $\mra>\mrb$,
and its stationary value becomes negative
at some point between 
the situation in the center of 
Fig.~\ref{cisaillement_evolution_contraintes_normales_k1_1_k2_0}
($\mra=\mrb=1/2$) and the situation on the right-hand side
($\mra=1$ and $\mrb=0$).

\subsection{Impact of plasticity}
\label{impact_of_plasticity}

Let us now briefly explore
the impact of the plastic deformation rate $\epspp(\Finger)$
through the influence of parameter $b_2$
in Eq.~(\ref{epspp_decomp_Finger}).

\begin{figure*}
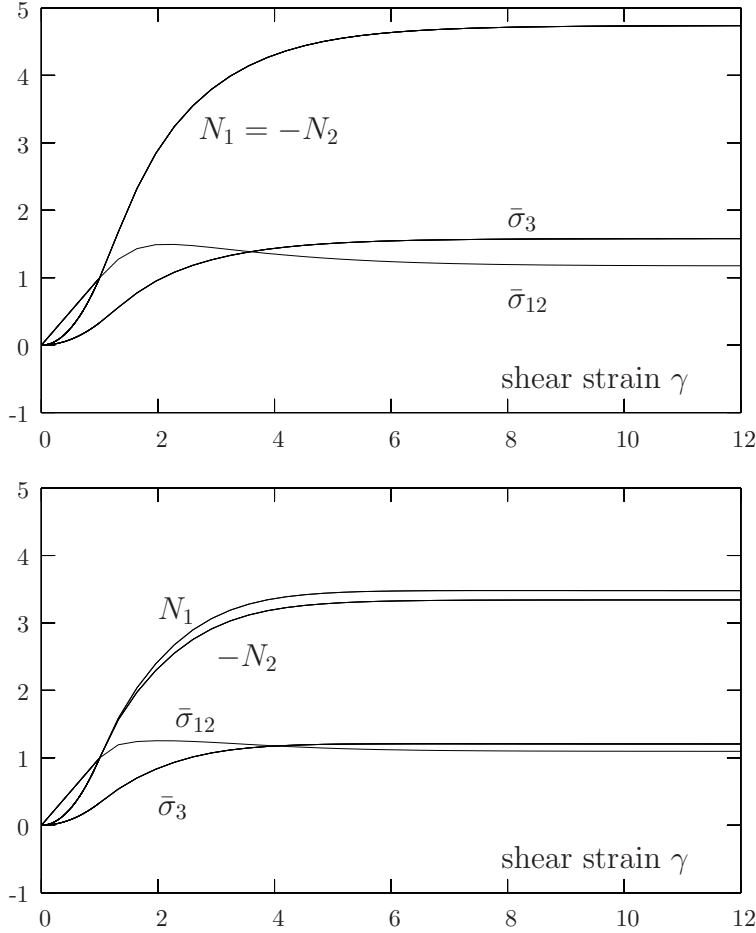

\begin{center}
\resizebox{0.65\columnwidth}{!}{%
  \input{cisaillement_evolution_contraintes_normales_k1_0_k2_1.pstex_t}
}
\resizebox{0.65\columnwidth}{!}{%
  \input{cisaillement_evolution_contraintes_normales_k1_0_k2_1_b2_egale_b1.pstex_t}
}
\end{center}
\caption{Impact of plasticity on the stress response.
The shear stress $\sud$, 
the deviatoric stress $\strois$ in the vorticity direction
and the normal stress differences $N_1$ and $N_2$
are plotted as a function of the shear deformation $\gamma$,
for a shear rate $\gd=8$.
Mooney-Rivlin elasticity is chosen with $k_1=0$ and $k_2=1$.
The plastic deformation rate is taken
as expressed by Eq.~(\ref{epspp_decomp_Finger})
with $b_1=(\trace\Finger-4)\theta(\trace\Finger-4)$.
Parameter $b_2$ is varied.
Left: $b_2=0$. Right: $b_2=b_1$.}
\label{cisaillement_evolution_contraintes_normales_k1_0_k2_1_b2_egale_b1}
\end{figure*}

In Fig.~\ref{cisaillement_evolution_contraintes_normales_k1_0_k2_1_b2_egale_b1},
we take $b_1=(\trace\Finger-4)\theta(\trace\Finger-4)$
like in Figs.~\ref{cisaillement_evolution_s12_k1_17_k2_67}--%
\ref{cisaillement_permanent_contraintes_normales_k1_17_k2_67}.
We then observe the (unspectacular) effect
of choosing $b_2=b_1$ rather than $b_2=0$.
Note that in this case,
as opposed to the role of elasticity
in paragraph~\ref{impact_of_elasticity},
the time evolution of tensor $\Finger$
is affected by the choice of the plastic law.

\subsection{Discussion}

In paragraphs~\ref{impact_of_elasticity}
and~\ref{impact_of_plasticity} above,
we illustrated the fact 
that the stress response of the material
can be altered through a change
in the material elasticity and plasticity alone.

Our approach emphasizes the fact that the choice
of the convective derivative
is somewhat arbitrary.
A material is often considered
as rather ``upper-convected''
or rather ``lower-convected''
because its stress evolution
follows more closely the eponymous convective derivative.
In fact, always choosing the upper-convected derivative
(and tensor $\Finger$ for a measure of the deformation)
would be equivalent, 
provided the elastic and plastic laws be suitably adjusted.
A forthcoming paper~\cite{discussion_on_derivatives}
will discuss in detail the consistency
of commonly used rheological laws
(see Section~\ref{consistency_stress_evolution} above)
and explore these issues more in depth.

\section{Conclusion and perspectives}
\label{conclusion_and_perspectives}

In order to describe the rheological behaviour
of foams and emulsions (and potentially other materials),
we have developped a continuous framework
to describe the evolution of the stored deformation tensor.
It includes elasticity, 
up to the large deformations
commonly encountered in such systems,
and plasticity.

Within this framework, we showed that it is possible
to play with various expressions
for the elasticity and the plasticity.
We hope that it is thus possible 
to adequately describe the rheological behaviour
of a large range of foams and emulsions
with dispersed phase volume fraction close to unity.

In less concentrated foams and emulsions,
volume fraction should be added as an extra field
to account for osmotic compressibility of the dispersed phase.
The issue whether the model may be adapted
to describe usual, aging complex fluids,
is left for future investigations.

\begin{figure*}
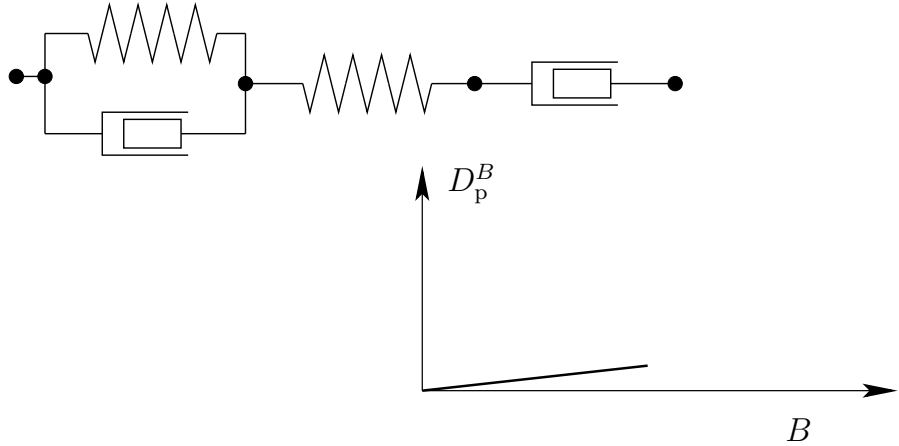

\begin{center}
\resizebox{0.55\columnwidth}{!}{%
  \input{modele_burger.pstex_t}
}
\resizebox{0.5\columnwidth}{!}{%
  \includegraphics{vide.eps}
}
\resizebox{0.4\columnwidth}{!}{%
  \input{epspp_fonction_de_Finger_burger.pstex_t}
}
\end{center}
\caption{Burger model: 
schematic diagramme (left)
and plot (right) of plastic flow 
as a function of tensor $\Finger$,
in the stationary regime and under small applied streses.
In Burger's model, suitable for foams
under low applied stresses 
(see paragraph~\ref{burger_model_for_weak_applied_stresses}),
the plastic flow increases with the applied stress.
If we neglect the short time scale response
provided by the Voigt-Kelvin element
(spring and viscous element in parallel),
then this increase is weak and linear,
as determined by the viscous element in series.}
\label{modele_bingham_burger_burger}
\end{figure*}

\begin{figure*}
\begin{center}
\resizebox{0.6\columnwidth}{!}{
  \includegraphics{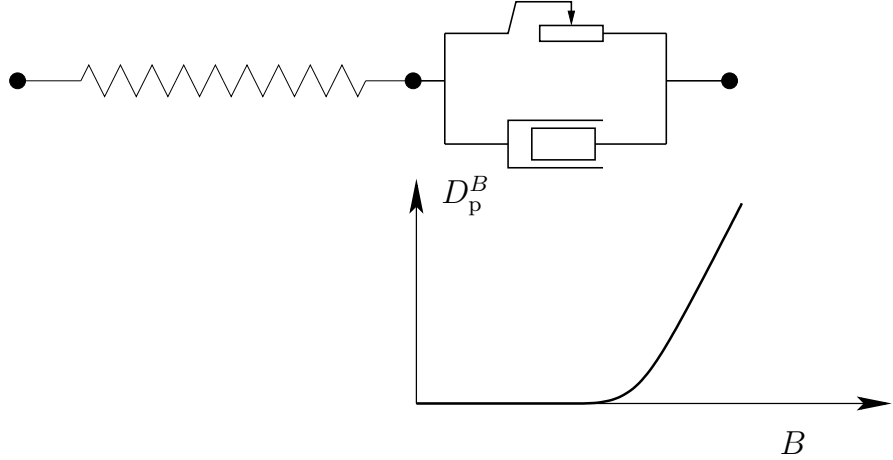}
}
\resizebox{0.45\columnwidth}{!}{%
  \includegraphics{vide.eps}
}
\resizebox{0.4\columnwidth}{!}{%
  \input{epspp_fonction_de_Finger_bingham.pstex_t}
}
\end{center}
\caption{Bingham model: 
schematic diagramme (left)
and plot (right) of plastic flow 
as a function of tensor $\Finger$,
in the stationary regime.
In a model based on Bingham's model,
such as that developed in the present work,
the plastic flow is zero up to a certain stress (yield stress).
Beyond the yield stress, it increases (linearly with the applied stress).}
\label{modele_bingham_burger_bingham}
\end{figure*}

\begin{figure*}
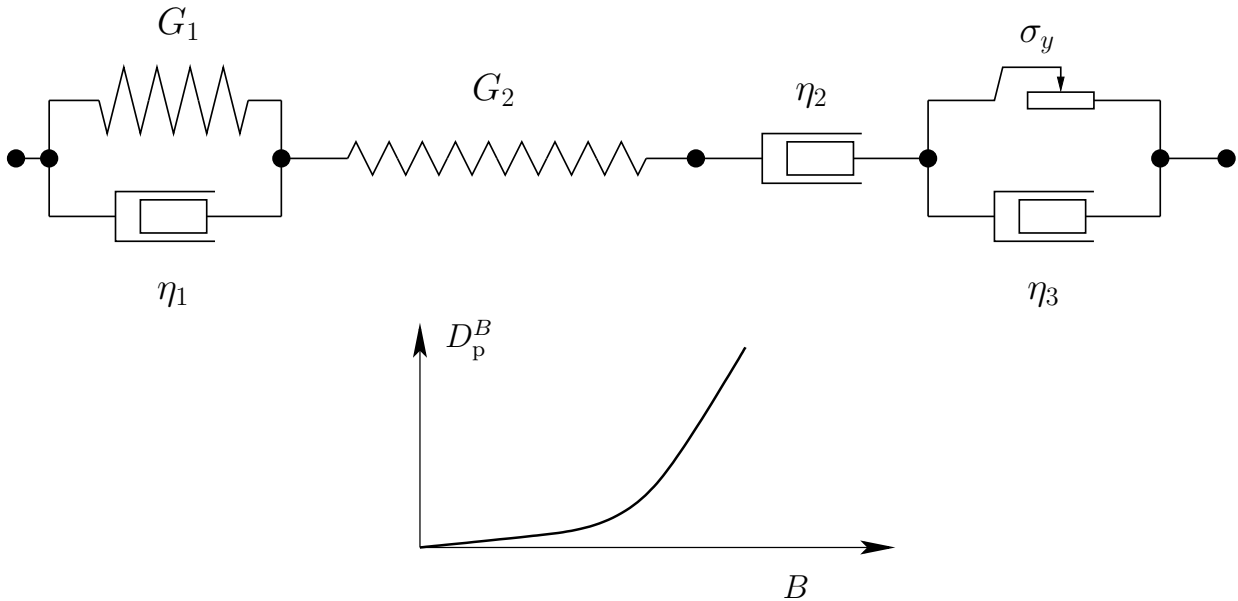

\begin{center}
\resizebox{1.0\columnwidth}{!}{%
  \input{modele_bingham_burger.pstex_t}
}
\resizebox{0.05\columnwidth}{!}{%
  \includegraphics{vide.eps}
}
\resizebox{0.4\columnwidth}{!}{%
  \input{epspp_fonction_de_Finger_burger_bingham.pstex_t}
}
\end{center}
\caption{Combined Bingham-Burger model: 
schematic diagramme (left)
and plot (right) of plastic flow 
as a function of tensor $\Finger$,
in the stationary regime.
In a combined Bingham-Burger model,
the plastic flow would increase weakly with the applied stress
at low stresses, and more strongly above the yield stress.
We believe it would mimic 
the rheological behaviour of a foam quite adequately.}
\label{modele_bingham_burger_bingham_burger}
\end{figure*}

As mentioned in paragraph~\ref{burger_model_for_weak_applied_stresses},
the Burger model (see Fig.~\ref{modele_bingham_burger_burger})
describes the response
of dry foams under weak stresses adequately.
By contrast, the Bingham model
(used as a basis for our formulation,
see Fig.~\ref{modele_bingham_burger_bingham})
captures the large stress behaviour.
It would be interesting to combine both models,
as represented in Fig.~\ref{modele_bingham_burger_bingham_burger}.
We did not elaborate on this combined model,
as it is technically more complex
than our Bingham-like model,
and we did not want to focus on short timescales
or weak stresses, where both models have differing behaviours.

\subsection{Determining the elasticity from experiments}

Beyond all these rheological and phenomenological models,
it is important to make the connection
with the microscopic scale.

From this point of view, the main issue
is the definition of the deformation in the material.
In the present work, deformation is built on a thought experiment:
we cut out a piece of material at sufficiently large a length scale
for disorder to be smoothed out.

By contrast, a way to define deformation was introduced 
by Aubouy {\em et al.}~\cite{grenoble_theo_2003}
for all systems in which the individual objects
and their mutual contacts are experimentally accessible,
such as 2D foams and emulsions 
(and 3D foams and emulsions when tomography
will have become a routine technique for imaging such systems
on rheologically relevant timescales).
They first construct a symmetric tensor $M$
from the centre-to-centre 
vectors~\cite{graner_dollet_raufaste_marmottant_2007}
for pairs of first neighbours.
This tensor is a dilation (proportional to the unit tensor
when the material is at rest).
They then define the deformation $U$ as the logarithm of $M$.

One might think that their tensor $U$ is just one deformation
among many others.
But with respect to the particular systems that they consider
(2D foams in the dry limit under quasistatic deformation),
it plays a very special role:
it is the deformation for which the elastic law
remains linear up to large 
deformations~\cite{grenoble_exp_2003}
(in fact, up to the onset of plasticity).
This result confers some weight to the initial,
scale-independence arguments~\cite{grenoble_theo_2003}
for the choice of the logarithm as the link
between their texture (dilation) tensor $M$
and their deformation measure $U=\log M$.

In our perspective, as we will show in a more elaborate manner
elsewhere~\cite{discussion_on_derivatives},
the choice of a deformation measure
corresponds to a choice for the convective derivative.
The deformation measure
$U$~\cite{grenoble_theo_2003,graner_dollet_raufaste_marmottant_2007}
could therefore lead directly to a continuum formalism
well-suited for (at least) 
two-dimensional foams in the elastic regime.

\subsection{Plasticity and mechanical noise}
\label{section_dissipation}

In the present work, the plastic deformation rate $\epspp$
was assumed to depend only on the local stress
{\em via} the local stored deformation,
see Eq.~(\ref{eq_elast_plast}).

This dependence is sufficient to account
for the non-local elastic effects
observed in foams~\cite{debregeas_tabuteau_di_meglio,kabla_debregeas},
and mentioned in paragraph~\ref{history_localization}.
Indeed, when plastic events occur (non-zero $\epspp$),
this impacts the local stored deformation
{\em via} Eq.~(\ref{dFingerplast}).
This, in turn, affects the stress
due to elasticity (Eq.~\ref{sidWdFinger}).
Equation (\ref{force_balance_equation})
then implies that the stress is modified
in the surrounding material.

Thus, within this continuum model
in which the plastic deformation rate
depends solely on the stored deformation (or stress),
plasticity at one location
alters the stress (and the stored deformation)
elsewhere in the material,
thus possibly contributing to triggering plasticity there.

Nevertheless, stress may not be the only factor
that determines the rate at which $T1$ processes occur.
For instance, in a recent work by
Marmottant and Graner~\cite{marmottant_graner},
the plastic deformation rate ($\epspp$ in our notation) 
is proportional to the total deformation rate $\epsp$
when it has the same sign as the stored 
deformation\footnote{a formulation specific to 2D materials.}.
Incidentally, such a choice implies that 
relaxation\footnote{reduction of the stored deformation
while the total deformation rate is zero.}
cannot take place in the system.

The fact that $\epspp$ depends on $\epsp$
can be interpreted physically
as the fact that mechanical noise may well help 
triggering plastic events.

As an alternative implementation of the impact of mechanical noise,
one might take the plastic deformation rate $\epspp$
as slightly enhanced in the presence
of non-zero total deformation rate
(although mainly determined by the stored deformation).
For instance, if we assume that the mechanical noise is isotropic,
we may include a multiplicative factor
of the form $(1+|\epsp^2|\,\tau^2)$ into $\epspp$,
where plasticity would be significantly enhanced
for deformation rates around and above $\tau^{-1}$.

We did not enter such subtelties in the present work,
and restricted our study to a purely stress-dependent
plastic deformation rate, 
as expressed by Eq.~(\ref{eq_elast_plast}).


\subsection{Plasticity and fluidity or fragility}
\label{section_fluidity}

In the present model,
the evolution of the stored deformation is given
by Eq.~(\ref{dFingerplast}):
$$
\frac{{\rm d}\Finger}{{\rm d}t}
-\gradv \cdot \Finger
-\Finger \cdot \transp{\gradv}
=-2\,\epspp
$$

In the present work, the plastic flow rate $\epspp$
was assumed to depend solely 
on the stored deformation $\Finger$:
\be
\epspp=\epspp(\Finger)
\ee
More generally, as mentioned 
in paragraph~\ref{section_dissipation} above,
one can include the effect of, say, mechanical noise
by including a dependence of $\epspp$
on the applied deformation rate $\epsp$:
\be
\label{epspp_fonction_Finger_et_epsp}
\epspp=\epspp(\Finger,\;\epsp)
\ee

In reality, the plastic flow rate $\epspp$
may well depend on yet other variables
than $\Finger$ and $\epsp$.

A few years ago, in their study of sheared
two-dimen\-sio\-nal foams between two solid plates,
Kabla and Debr\'egeas noticed that $T1$ events
appeared preferentially in regions
where the stress tensor was most disordered~\cite{kabla_debregeas}.

Very recently, Goyon and coworkers~\cite{goyon_fluidity}
studied the flow of three-dimensional emulsions
in milli-fluidic channels.
Because of the large aspect ratio of the channel section,
they were able to minimize edge effects
and to obtain two-dimensional velocity profiles
for the stationary flow of such emulsions between two walls.

One of their key results is that even though they are able
to measure locally the relation 
between the shear stress 
(which they derive from the applied pressure)
and the shear rate (which they observe),
they do not obtain a single mastercurve $\si=\si(\gd)$
when changing the applied pressure.

By contrast, Eq.~(\ref{dFingerplast}),
combined with Eq.~(\ref{epspp_fonction_Finger_et_epsp}),
would predict, in stationary flow,
that the stored deformation could be expressed
in terms of the velocity gradient,
$\Finger=\Finger(\gradv)$, 
and similarly for the stress, {\em via} elasticity.
Hence, in the present situation, they would predict:
$$
\si=\si(\gd)
$$

The observed behaviour~\cite{goyon_fluidity}
contradicts this prediction.
It thus shows that the plastic flow rate
must depend on additional variables.
The observations have been shown~\cite{goyon_fluidity}
to be compatible with the existence of diffusive, scalar quantity $\Gamma$,
which they call {\em fluidity}.

It will be a challenge, in future studies,
to identify the microscopic origin of such a quantity,
which should be truly tensorial in more general situations than plane shear.
In the quasistatic limit, 
fluidity should probably be related to the frozen stress disorder
identified by Kabla and Debr\'egeas~\cite{kabla_debregeas}.

\subsection{Liquid permeation}

\begin{figure}[ht!]
\begin{center}
\resizebox{0.4\columnwidth}{!}{\input{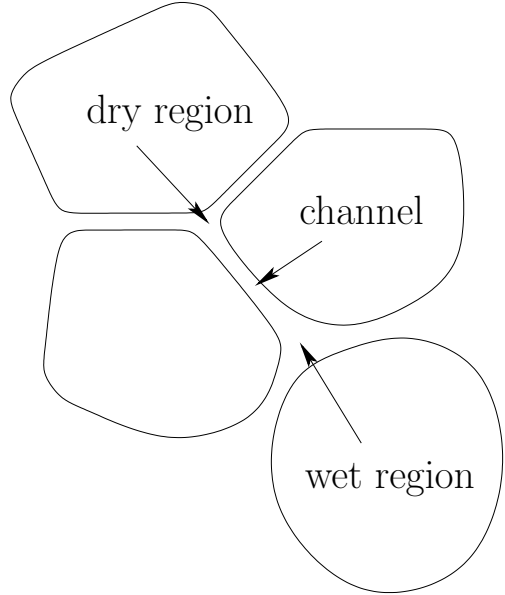}}
\end{center}
\caption{Foam with inhomogeneities in liquid volume fraction.
The gas/liquid interface is flat in inter-bubble films,
and curved in vertex regions at the junction 
between 3 (in two dimensions) or 4 (in three dimensions) bubbles.
This curvature implies that the liquid pressure in the vertices
is smaller than the gas pressure in the neighbouring bubbles.
In wet regions, the interface curvature is less pronounced
than in dry regions.
As a result, there exists a pressure gradient,
and the liquid tends to flow from wet regions towards dry regions.
The intensity of the liquid flow
depends on the hydraulic resistance
in the Plateau borders that convey most of the liquid
(depicted as the channel in this two-dimensional drawing).}
\label{pression_et_mousse_seche_humide}
\end{figure}

Permeation of the continuous, liquid phase
through the network of channels formed
by the bubbles or droplets
may result from gravity,
due to the density mismatch between both phases,
and is called {\em drainage}.
Even in the absence of gravity, permeation may occur 
when the volume fraction of both phases
is not uniform throughout the foam.
Indeed, as depicted on Fig.~\ref{pression_et_mousse_seche_humide},
the pressure in the liquid phase is lower
in dryer regions than in wetter regions
for otherwise identical pressure values in the gas bubbles.
As a result, the liquid permeates from wetter regions
towards dryer regions.

Since vector field $\vec{v}$
represents the {\em bubble} (or {\em droplet}) velocity
while the material density $\rho$
includes the mass of the {\em liquid phase}
(see paragraph~\ref{two_phase_fluid}),
the density conservation Eq.~(\ref{density_evolution})
must include a term that reflects permeation.

This effect can be incorporated into
Eq.~(\ref{density_evolution}) 
in a rough manner
by adding a diffusion term:
\be
\label{density_evolution_with_permeation}
\frac{\partial\rho}{\partial t}
+\nabla\cdot(\rho\;\vec{v})
=D_{\rm permeation}\,\Delta\rho
\ee
where the diffusion coefficient $D_{\rm permeation}$
depends on the geometrical dimensions 
of the Plateau borders between foam vertices
(which depend mainly on the volume fraction $\varphi$)
but also on the hydrodynamic boundary conditions
along the Plateau borders (which depend
on various preparation conditions
such as surfactant nature and concentration, salt, etc,
in a very non-trivial manner).

\subsection{Towards dilatancy}

Eq.~(\ref{density_evolution_with_permeation}) above
is only the first step towards a model
that would include other effects
known to exist in foams and emulsions.
Indeed, relaxing the assumptions
made in paragraph~\ref{evolution_modes}
would enable us to include an important phenomenon
which is well known in the context of granular media
and which has been recently demonstrated
in liquid foams~\cite{foam_dilatancy},
namely {\em dilatancy}.
When deformed, the local conformation of the foam, 
as schematized on Fig.~\ref{permeation_gas_diffusion_dilation},
tends to go from conformation $(0)$
to conformation $(4)$.
The physical origin of such a phenomenon is not clearly understood yet.
It might be related to the thickening of films~\cite{emile_2007}
or Plateau borders~\cite{emile_2007,terriac_2006}
observed when a Plateau border glides at a solid wall.

Together with the identification
of the microscopic origin of fluidity
(see paragraph~\ref{section_fluidity} above),
including the effect of dilatancy
will thus be yet another challenge
in the ever-bewildering rheology of foams and emulsions.

\section*{Acknowledgements}

We gratefully thank 
Miguel Aubouy, 
Annie Colin, 
François Graner, 
Jean-Charles Razafindrakoto,
Reinhard H\"ohler
and Philippe Marmottant
for stimulating discussions,
and Pierre Rognon for a critical reading of the manuscript.

\appendix

\section{Evolution of Finger tensor}
\label{appendix_dBdt}

\begin{figure}[ht!]
\begin{center}
\resizebox{0.9\columnwidth}{!}{\input{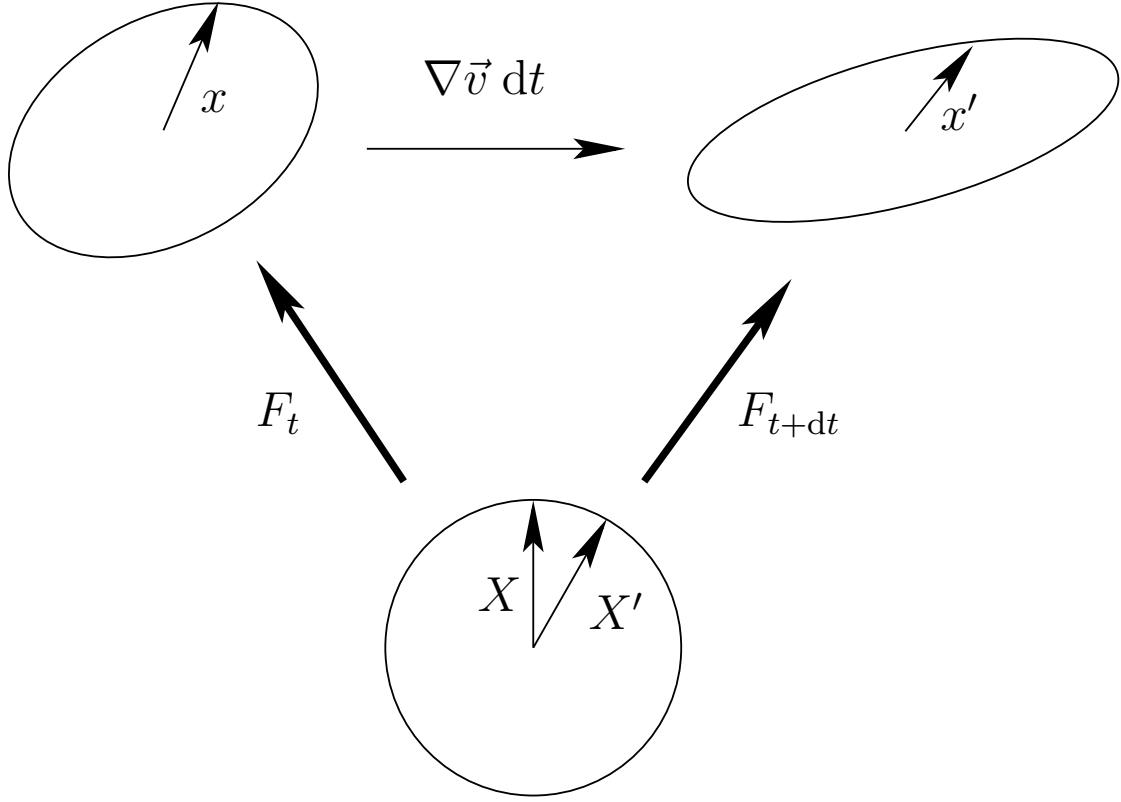}}
\end{center}
\caption{Augmented version of Figure~\ref{dSigma_plus_dEpsilon}.
As mentioned in the caption of Figure~\ref{decoupage},
the relaxed state local orientation is chosen in such a way
that tensor $\FF$ is a pure deformation.
As a consequence, points $X$ and $X^\prime$
do not coincide except for particular values of $\gradv$.}
\label{dSigma_plus_dEpsilon_avec_prime}
\end{figure}

In order to derive the evolution 
of tensor $\Finger$ (see Eq.~\ref{Finger_beta_i})
in some known velocity field,
we compute how an ellipse with generic point $x$
(see Eq.~\ref{eq_ellipsoid})
is convected into a new ellipse
(see Fig.~\ref{dSigma_plus_dEpsilon}).
The generic point $\vxp$ of the new ellipse
is obtained as:
\be
\vxp=(1+\gradv\;{\rm d}t)\cdot\vx
\ee
The equation for the new ellipse can be written in two ways:
\bee
\label{eq_ellipsoid_dt_1}
&\vxpt\cdot\left[\FF^{-2}\right]_{t+{\rm d}t}\cdot\vxp=R^2&\nonumber\\
\label{eq_ellipsoid_dt_2}
&\vxpt\cdot(1-\transp{\gradv}\;{\rm d}t)\cdot\left[\FF^{-2}\right]_{t}\cdot(1-\gradv\;{\rm d}t)\cdot\vxp=R^2&\nonumber
\eee
The evolution of tensor $\Finger^{-1}=\FF^{-2}$ is thus given by:
\be
\label{dFF}
\frac{{\rm d}}{{\rm d}t}
\left[\Finger^{-1}\right]
=-\transp{\gradv} \cdot \Finger^{-1}
- \Finger^{-1} \cdot \gradv
\ee
{From} Eq.~(\ref{dFF}) above,
we then compute\footnote{To derive Eq.~(\ref{dFingerdt})
{from} Eq.~(\ref{dFF}), we use the identity:
$$
0={\rm d}[\Finger\cdot\Finger^{-1}]/{\rm d}t
=\Finger\cdot{\rm d}[\Finger^{-1}]/{\rm d}t
+[{\rm d}\Finger/{\rm d}t] \cdot\Finger^{-1}
$$}
the evolution of tensor $\Finger$,
see Eq.~(\ref{dFingerdt}).


\section{Density and stored deformation}
\label{density_and_stored_deformation}

Our assumptions stated in paragraph~\ref{evolution_modes}
imply that the material density
is related to the determinant of tensor $\Finger$
in a very simply way.
The evolution of $\det\Finger$ can always be written in the form:
\be
\label{identity_evol_det_Finger}
\frac{{\rm d}(\det\Finger)}{{\rm d}t}
=\trace\left[
(\det\Finger)\,\Finger^{-1}\cdot
\frac{{\rm d}\Finger}{{\rm d}t}
\right]
\ee

After multiplying Eq.~(\ref{dFingerplast})
by $\inv{\Finger}$, taking the trace
and using the plastic incompressibility
expressed by Eq.~(\ref{incompressibilite_plastique}),
we insert the result into Eq.~(\ref{identity_evol_det_Finger})
and obtain:
\be
\label{evol_det_Finger}
\frac{{\rm d}(\det\Finger)}{{\rm d}t}
=2(\det\Finger)\;(\trace\epsp)
\ee
and finally:
\be
\label{evol_sqrt_det_Finger}
\frac{{\rm d}(1/\sqrt{\det\Finger})}{\dt}
+(1/\sqrt{\det\Finger})\;(\trace\epsp)
=0
\ee
Comparing Eqs.~(\ref{density_evolution})
and~(\ref{evol_sqrt_det_Finger})
shows that for any given element of material,
the current density and deformation
are linked {\em via}
their initial values:
\be
\frac{\rho(t,\vec{r})}{\rho(t_0,\vec{r_0})}
=\frac{\sqrt{\det\Finger(t_0,\vec{r_0})}}
{\sqrt{\det\Finger(t,\vec{r})}}
\ee
where $\vec{r_0}$ is the position
of material point $\vec{r}$ at time $t_0$.


\section{Shear flow}
\label{calculs}

\subsection{Derivation of the shear flow equations}
\label{derivation}

In order to derive the evolution of tensor $\Finger$
(Eqs.~\ref{cos2theta}--\ref{dcos2thetadt}),
let us use the notations of Fig.~\ref{xyXYz}.

The shear velocity gradient, the material deformation $\Finger$
and the flow rate $\epspp$ have the following form
in basis $xyz$:
\bee
\label{gradv_form_xyz}
\gradv^{[xyz]}&=&
\begin{pmatrix}0 & \gd & 0 \\
0&0&0\\ 
0&0&0
\end{pmatrix}\\
\label{B_form_xyz}
\Finger^{[xyz]}&=&
\begin{pmatrix}
\ct^2\beta_1+\st^2\beta_2 & \ct\st(\beta_1-\beta_2) & 0 \\
\ct\st(\beta_1-\beta_2) & \st^2\beta_1+\ct^2\beta_2 &0\\ 0&0& \beta_3
\end{pmatrix}\\
\label{Dp_form_xyz}
\epspp^{[xyz]}&=&\begin{pmatrix}
\ct^2\delta_1+\st^2\delta_2 & \ct\st(\delta_1-\delta_2) & 0 \\
\ct\st(\delta_1-\delta_2) & \st^2\delta_1+\ct^2\delta_2 &0\\ 0&0& \delta_3
\end{pmatrix}
\eee
where $s$ (resp., $c$) denotes the sine (resp., the cosine)
of angle $\theta$ between axes $x$ and $X$
(see Fig.~\ref{xyXYz}).

Inserting the above equations into the evolution equation~(\ref{dFingerplast}),
we obtain:

\bee
\label{equadif1}
\dot{\overbrace{\ct^2\beta_1+\st^2\beta_2}}&=&2\gd\ct\st(\beta_1-\beta_2)-2(\ct^2\delta_1+\st^2\delta_2)\\
\label{equadif2}
\dot{\overbrace{\ct\st(\beta_1-\beta_2)}}&=&\gd(\st^2\beta_1+\ct^2\beta_2)-2\ct\st(\delta_1-\delta_2)\\
\label{equadif3}
\dot{\overbrace{\st^2\beta_1+\ct^2\beta_2}}&=&-2(\st^2\delta_1+\ct^2\delta_2)
\eee
Summing Eqs.~(\ref{equadif1}) and~(\ref{equadif3})
yields the evolution of $\beta_1+\beta_2$.
The difference between Eqs.~(\ref{equadif1}) and~(\ref{equadif3}),
multiplied by $(\beta_1-\beta_2)\cos(2\theta)$,
plus Eq.~(\ref{equadif2}) multiplied by $2(\beta_1-\beta_2)\sin(2\theta)$,
yields the evolution of $\beta_1-\beta_2$.
From there, the evolution of $\beta_1$ and $\beta_2$
is readily obtained, see Eqs.~(\ref{dbeta1dt}) and~(\ref{dbeta2dt}).
Using these equations,
the difference between Eqs.~(\ref{equadif1}) and~(\ref{equadif3})
then yields the evolution of $\cos(2\theta)$,
see Eq.~(\ref{dcos2thetadt})

\subsection{Elastic shear flow}
\label{elasticshearflow}

Let the initial configuration be a material at rest,
with $\beta_1=\beta_2=1$.
The system of Eqs.~(\ref{dbeta1dt}), (\ref{dbeta2dt})
and~(\ref{dcos2thetadt}) has a singularity at $t=0$
since Eq.~(\ref{dcos2thetadt}) contains a factor $\frac{1}{\beta_1-\beta_2}$.
In fact, the system of Eqs.~(\ref{equadif1}), (\ref{equadif2}) and~(\ref{equadif3})
can be solved explicitely in the domain
where the material is purely elastic, with $\delta_1=\delta_2=0$.
Using the same notation as in Appendix~\ref{derivation}:
\bee
\label{interm1}
\ct^2\beta_1+\st^2\beta_2&=&1+\gd^2t^2\\
\label{interm2}
\ct\st(\beta_1-\beta_2)&=&\gd t\\
\label{interm3}
\st^2\beta_1+\ct^2\beta_2&=&1
\eee
Whence:
\bee
\gamma(t)&=&\gd\;t\\
\label{beta1el}
\beta_1(t)&=&1+\frac{\gamma(t)^2}{2}+\frac{\gamma(t)\;\sqrt{\gamma(t)^2+4}}{2}\\
\label{beta2el}
\beta_2(t)&=&1+\frac{\gamma(t)^2}{2}-\frac{\gamma(t)\;\sqrt{\gamma(t)^2+4}}{2}\\
\label{uel}
u(t)&=&\cos(2\theta)=\frac{\gamma(t)}{\sqrt{\gamma(t)^2+4}}
\eee

\subsection{Plasticity onset}
\label{plasticity_onset}

When in the elastic regime,
the system is described by Eqs.(...) above.
Once the threshold is reached,
given by Eq.~(\ref{threshold_function}),
the plastic term comes into play.

Eq.~(\ref{threshold_function}) by itself
describes the limit of the elastic regime
in terms of tensor $\Finger$
for any choice of the system history in the elasic regime
(not just the simple shear implemented here).
It is represented 
on Fig.~\ref{cisaillement_evolution_betai_k1_17_k2_67}.

\subsection{Stationary shear flow}
\label{stationaryshearflow}

When the flow is stationary,
the system of Eqs~(\ref{dbeta1dt}--\ref{dcos2thetadt})
can be simplified.
In particular, Eq.~(\ref{dcos2thetadt}) then implies:
\bee
\cos(2\theta)&=&
u=\frac{\beta_1-\beta_2}{\beta_1+\beta_2}\\
\label{racinede1moinsu2}
\sin(2\theta)&=&
\sqrt{1-u^2}=\frac{2\sqrt{\beta_1\beta_2}}{\beta_1+\beta_2}
\eee
and finally the tilt $\theta$
of basis $XY$ relative to basis $xy$
can be expressed as:
\be
\theta=\arctan\sqrt{\frac{\beta_2}{\beta_1}}
\ee

From Eqs~(\ref{dbeta1dt}) and~(\ref{dbeta2dt}),
still in the stationary regime, we obtain:
\be
\label{delta_3_egale_0}
\frac{\delta_1}{\beta_1}
+\frac{\delta_2}{\beta_2}=0
\ee
Combining this with Eq.~(\ref{incompressibilite_plastique})
which expresses the material incompressibility,
we obtain $\delta_3=0$.

Again from Eqs~(\ref{dbeta1dt}) and~(\ref{dbeta2dt})
in the stationary regime,
we derive the shear rate:
\bee
\gd\;\sqrt{1-u^2}
&=&\frac{2\delta_1}{\beta_1}
=-\frac{2\delta_2}{\beta_2}\nonumber\\
&=&\frac{2\delta_1-2\delta_2}{\beta_1+\beta_2}
\eee
Using Eq.~(\ref{racinede1moinsu2}), we then obtain:
\be
\label{gdfrombetaianddeltai}
\gd=\frac{\delta_1-\delta_2}{\sqrt{\beta_1\beta_2}}
\ee

Thus, the stationary branch of the flow plots
are obtained by choosing pairs of values 
for $\beta_1$ and $\beta_2$
that satisfy Eq.~(\ref{delta_3_egale_0}),
where $\delta_1$ and $\delta_2$
are functions of $\beta_1$, $\beta_2$
and $\beta_3=(\beta_1\beta_2)^{-1}$.
To obtain the relevant values for $\beta_1$ and $\beta_2$,
we first find the pair that verifies simulataneously
the threshold condition given by Eq.~(\ref{threshold_function})
and the stationary condition given by Eq.~(\ref{delta_3_egale_0}).
We then use a differential equation
derived from Eq.~(\ref{delta_3_egale_0})
to follow the corresponding curve in the $\beta_1$--$\beta_2$ plane.

Once the values for $\beta_1$ and $\beta_2$
are known, the shear rate is derived
{\em via} Eq.~(\ref{gdfrombetaianddeltai}).\newline

We can now slightly simplify 
Eqs.~(\ref{s_xyz})--(\ref{N2_MR_shear}) 
in the stationary regime.

\bee
\label{s_xyz_stationary}
\s&=&\begin{pmatrix}
\frac{\beta_1\su+\beta_2\sd}{\beta_1+\beta_2}
&\frac{\sqrt{\beta_1\beta_2}(\su-\sd)}{\beta_1+\beta_2}
&0\\
\frac{\sqrt{\beta_1\beta_2}(\su-\sd)}{\beta_1+\beta_2}
&\frac{\beta_1\sd+\beta_2\su}{\beta_1+\beta_2}
&0\\ 0&0&\strois
\end{pmatrix}
\eee
The first and second normal stress difference can be expressed as:
\bee
N_1&\equiv&\suu-\sdd\nonumber\\
&=&\frac{\beta_1-\beta_2}{\beta_1+\beta_2}(\su-\sd)\nonumber\\
&=&\frac{(\beta_1-\beta_2)^2}{\beta_1+\beta_2}
\left[a_1+a_2(\beta_1+\beta_2)\right]\\
N_2&\equiv&\sdd-\strois\nonumber\\
&=&\frac{\beta_1(\sd-\strois)+\beta_2(\su-\strois)}{\beta_1+\beta_2}\nonumber\\
&=&a_1\left[2\frac{\beta_1\beta_2}{\beta_1+\beta_2}
-\frac{1}{\beta_1\beta_2}\right]\nonumber\\
&&+a_2\left(\beta_1\beta_2-\frac{1}{(\beta_1\beta_2)^2}\right)
\eee
In the case of Mooney-Rivlin,
coefficients $a_1$ and $a_2$ are given
by Eq.~(\ref{a1_a2_mooney_rivlin})
and the normal stress differences
can be expressed as:
\bee
N_1&=&\frac{(\beta_1-\beta_2)^2}{\beta_1+\beta_2}
\left[\mra+\frac{\mrb}{\beta_1\beta_2}\right]\\
N_2&=&\mra\left[\frac{2\beta_1\beta_2}{\beta_1+\beta_2}
-\frac{1}{\beta_1\beta_2}\right]\nonumber\\
&&+\mrb\left[\beta_1\beta_2
-\frac{\beta_1+\beta_2}{\beta_1\beta_2}
+\frac{2}{\beta_1+\beta_2}\right]
\eee


%
%
%
%
 \bibliographystyle{unsrt}
 \bibliography{fluche}

\end{document}